\documentclass[11pt]{article}
\usepackage[dvips]{epsfig}
\usepackage{graphicx}
\def\beq{\begin{eqnarray}}
\def\eeq{\end{eqnarray}}

\begin{document}

\fontsize{11}{14.5pt}\selectfont

\begin{center}

{\small Technical Report No.\ 9910 (revised),
 Department of Statistics, University of Toronto}

\vspace*{1in}

{\LARGE \bf Circularly-Coupled Markov Chain Sampling } \\[16pt]

{\Large Radford M. Neal}\\[3pt]
 Department of Statistics and Department of Computer Science \\
 University of Toronto, Toronto, Ontario, Canada \\
 \texttt{http://www.cs.utoronto.ca/\~{}radford/} \\
 \texttt{radford@stat.utoronto.ca}\\[10pt]

 \begin{tabular}{rl} First version: & 22 November 1999 \\
                     Revised:       & 28 February 2002
 \end{tabular}
\end{center}

\vspace{12pt} 

\begin{center} \bf Abstract \end{center}

I show how to run an $N$-time-step Markov chain simulation in a
circular fashion, so that the state at time $0$ follows the state at
time $N\!-\!1$ in the same way as states at times $t$ follow those at
times $t\!-\!1$ for $0\!<\!t\!<\!N$.  This wrap-around of the chain is
achieved using a coupling procedure, and produces states that all have
close to the equilibrium distribution of the Markov chain, under the
assumption that coupled chains are likely to coalesce in less than
$N/2$ iterations.  This procedure therefore automatically eliminates
the initial portion of the chain that would otherwise need to be
discarded to get good estimates of equilibrium averages.  The
assumption of rapid coalescence can be tested using auxiliary chains
started at times spaced between $0$ and $N$.  When multiple processors
are available, such auxiliary chains can be simulated in parallel, and
pieced together to give the circularly-coupled chain, in less time
than a sequential simulation would have taken, provided that
coalescence is indeed rapid.

The practical utility of these procedures is dependent on the
development of good coupling schemes.  I show how a specialized
``random-grid'' Metropolis algorithm can be used to produce the
required exact coalescence.  On its own, this method is not efficient
in high dimensions, but it can be used to produce exact coalescence
once other methods have brought the coupled chains close together.  I
investigate how well this combined scheme works with standard
Metropolis, Langevin, and Gibbs sampling updates.  Using such
strategies, I show that circular coupling can work effectively in a
Bayesian logistic regression problem.

\newpage

\section{Introduction}\label{sec-intro}\vspace*{-10pt}

Sampling from a complex distribution by simulating a Markov chain
having this distribution as its equilibrium distribution is an
important technique in statistical mechanics (eg, Frenkel and Smit
1996), in Bayesian statistics (eg, Gilks, \textit{et al} 1996), and in
other computational problems (eg, Sinclair 1993).  States from the
equilibrium distribution of the chain are then used to estimate
quantities of interest, such as the average energy of a physical
system, or a Bayesian predictive distribution.

Ideally, such simulations would be conducted with theoretical
knowledge of the time needed for the chain to reach its
equilibrium distribution to within a given tolerance.  Although some
progress has been made at producing quantitative bounds on convergence
times of Markov chains (eg, Rosenthal 1995a), such theoretical
guarantees are presently unavailable for most problems of practical
interest.

Instead, practitioners usually assess the convergence of the Markov
chain sampler empirically, by formal or informal methods.  These
methods attempt to determine whether the chain has reached an adequate
approximation to its equilibrium distribution within the number of
iterations simulated.  If it appears to have done so, some initial
portion of the chain (the ``burn-in'' period) is generally discarded
in order to avoid biasing the results by inclusion of states that
reflect the state in which the chain was started rather than its
equilibrium distribution.  The bewildering variety of methods for
diagnosing convergence and discarding an appropriate burn-in period
have been reviewed by Cowles and Carlin (1996), Brooks and Roberts
(1998), and Mengersen, \textit{et al} (1999).

One convergence diagnostic, due to Johnson (1996), looks at multiple
``coupled'' chains that are started from different initial states, but
that subsequently undergo transitions determined by the same random
numbers.  Rosenthal (1995b) reviews the application of coupling to
Markov chains.  Briefly, coupling chains introduces dependencies
between them, and may lead them to ``coalesce'' to the same state
after some number of iterations.  The probability that a chain started
from the initial state distribution has not coalesced with a chain
started from the equilibrium distribution by time $T$ provides an
upper bound on the total variation distance of the chain from
equilibrium at time $T$.  In Johnson's diagnostic, the time required
for several chains whose initial states were drawn from a distribution
that is meant to be ``overdispersed'' with respect to the equilibrium
distribution is taken as an informal indication of how much time is
required for approximate equilibrium to be reached.

One way of viewing the circular coupling method of this paper is as a
refinement of Johnson's scheme, which addresses two problems that
scheme suffers from.  One problem noted by Johnson is that using the
states immediately following the time when all chains coalesce
introduces a bias in the results, favouring states where coalescence
is more likely.  The circular coupling scheme of this paper discards
an initial portion of the chain without introducing bias (under
certain assumptions), by using the last state of the chain to start a
new chain at time zero, and using the states of this chain rather than
of the original chain up to the point where it and the original chain
coalesce.  Another problem is that although Johnson's scheme considers
several initial states, it uses only a single sequence of random
numbers.  It could be that this one sequence happens to produce
atypically fast coalescence.  Diagnostics in the circular coupling
scheme are based on a variety of starting states at times spaced
throughout the total simulation period, thereby effectively
considering various initial random number sequences.  This reduces
(but does not eliminate) the possibility that the results will be
misleadingly optimistic.

The coupling technique of this paper also provides a way of exploiting
parallel computation in Markov chain simulation.  As discussed by
Rosenthal (1999), there are many possibilities for exploiting multiple
processors for the overall task of estimating quantities using Markov
chain Monte Carlo methods.  However, the core operation of simulating
a single realization of a Markov chain might appear to be inherently
sequential, since it might seem that the state at time $t$ cannot be
obtained until the state at time $t\!-\!1$ has been found.  In this
paper, however, I will show how coupling allows one to use multiple
processors to simulate a single realization of a Markov chain in
substantially less time that would be required by a single processor,
provided that the Markov chain and the coupling method employed lead
to rapid coalescence of chains.

The practical feasibility of circular coupling is dependent on finding
an efficient coupling scheme --- ie, a way of introducing dependencies
between the transitions from chains currently in different states that
promotes the rapid coalescence of these chains to the same state.  In
this paper, I first describe a specialized ``random-grid'' Metropolis
update, which can produce exact coalescence even for continuous
distributions, and use this to demonstrate circular coupling in simple
one-dimensional examples.  In higher dimensions, this method does not
work well on its own.  However, one can combine more standard Markov
chain updates, coupled in a way that leads the chains to approach each
other more and more closely, with an occasional random-grid Metropolis
update, which can produce exact coalescence once the chains are very
close.  I examine how well this strategy works using standard
Metropolis, Langevin, and Gibbs sampling updates, and find that at
least the Langevin and Gibbs sampling updates can produce efficient
coupling.

I conclude by demonstrating that circular coupling can be applied
effectively to a polytomous logistic regression problem with a
hierarchical prior over regression coefficients.

Although circular coupling works well in this demonstration, there may
sometimes be a cost to restricting the Markov chains used to those for
which good coupling methods are available, and even when a suitable
coupling scheme is available, it may not be optimal, and hence may
lead to coalescence times that are greater than the actual time
required for the Markov chain to reach approximate equilibrium.
However, this requirement for an efficient coupling scheme is less
onerous than for the alternative of exact (a.k.a.\ ``perfect'')
sampling methods, such as coupling from the past (Propp and Wilson
1996) and the interruptible scheme of Fill (1998).  For these methods,
the coupling scheme must not only promote coalescence, but also permit
the efficient tracking of large sets of states, so that the
coalescence of a huge (possibly infinite) set of chains started in all
possible states can be determined.  Circular coupling looks only at
the coalescence of a moderate number of explicitly simulated chains,
and is therefore much easier.  The price of this is that circular
coupling will not provide an absolute guarantee that the states
obtained are from the exact equilibrium distribution, but only an
assurance that they are from close to the equilibrium distribution,
provided that certain conditions are met whose truth can be
empirically tested, but not verified with certainty.

In some situations, any added cost from using Markov chain methods
that produce good coupling may be offset by using this coupling to
reduce the variance of estimates.  This can be done by using
estimators that exploit correlations with states of a coupled chain
that samples from an approximating distribution for which expectations
are known exactly (Pinto and Neal 2001; Schmeiser and Chen 1991; Chen,
\textit{et al} 2000, Section 3.4).

I begin in Section~\ref{sec-alg} by presenting the algorithms and
their justification in the abstract.  Section~\ref{sec-rg} introduces
the random-grid Metropolis coupling scheme, uses it to demonstrate the
idea of circular coupling in two one-dimensional problems, and
discusses how this scheme can be generalized to higher-dimensional
problems, though on its own it does not perform well in a
nine-dimensional example.  Section~\ref{sec-schemes} shows how
random-grid Metropolis updates can be combined with other updates to
produce more efficient circularly-coupled samplers.  These methods are
shown to be effective in a logistic regression problem in
Section~\ref{sec-lr}.  I conclude in Section~\ref{sec-disc} by
discussing the significance of circular coupling for the routine use
of Markov chain sampling.

\section{Circular coupling algorithms}\label{sec-alg}\vspace*{-10pt}

Suppose we wish to sample from a distribution $\pi$ for some state
variable $x$ by using a Markov chain having $\pi$ as an invariant
distribution.  We hope that this Markov chain is ergodic, and hence
has $\pi$ as its only invariant distribution, and that it converges to
this equilibrium distribution rapidly.

Realizations of this chain with different initial states can be
coupled by representing the transitions of the chain by a function
$\phi(x_t,u_t)$, which takes as arguments the state at some time,
$x_t$, and the random numbers generated at that time, $u_t$, and
returns the state of the chain at the next time, $x_{t+1}$.  The
random numbers at each time are drawn independently, each from the
same distribution, $U$. 

With this representation, an ordinary Markov chain simulation for $N$
time steps is conducted as follows:
\begin{list}{}%
{\setlength{\leftmargin}{0.65in}%
\setlength{\labelwidth}{0.25in}%
\setlength{\topsep}{4pt}}
\item[\textbf{Standard Markov chain simulation:}\hfill]
\item[1)\hfill] Randomly draw $x_0$ from the initial state distribution, $p_0$.
\item[2)\hfill] For $t = 1,\ldots,N$: \\[4pt]
                \hspace*{0.2in}Randomly draw $u_{t-1}$ from the 
                distribution $U$, independently of previous draws. \\
                \hspace*{0.2in}Let $x_t = \phi(x_{t-1},u_{t-1})$.
\end{list}
There are many ways of expressing a given set of transition
probabilities by using different transition functions,
$\phi$, and different random number distributions, $U$.  Apart from
possible differences in computational cost, the choice makes no difference 
for the algorithm above.  

In the circular-coupling algorithms below, however, chains started in
different states are coupled by using the same $u_t$, in the hope that
this will lead the chains to ``coalesce'' to the same state.
Different ways of expressing the transitions in terms of a $\phi$
function may lead to coalescence occurring more or less rapidly.
However, with any coupling scheme of this nature, once two chains have
coalesced to the same state at some time, they will remain in the same
state at all subsequent times.

\subsection{The basic circular coupling procedure}\label{sec-proc}\vspace*{-6pt}

\begin{figure}
\centerline{\psfig{file=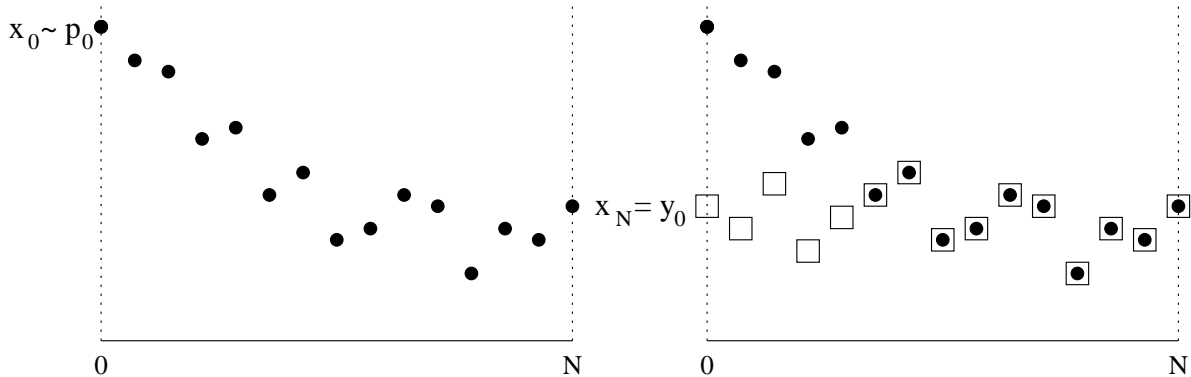,width=6.2in}}
\caption[]{The basic circular coupling procedure.  In this illustration,
the state is plotted on the vertical axis, and simulation time
is on the horizontal axis.  The plot on the left shows the generation of
the original chain, $x_0,\ldots,x_N$, starting with a state drawn from the
distribution $p_0$.  The plot on the right shows the
subsequent generation of the wrapped-around chain, $y_0,\ldots,y_N$.  In this 
case, the two chains coalesce at $t=5$.}\label{fig-basic}
\end{figure}

A circularly-coupled Markov chain simulation for $N$ time steps begins
the same way as a standard simulation, but after generating
$x_0,\ldots,x_N$, a second set of states, $y_0,\ldots,y_N$, are
generated by letting $y_0=x_N$ and then redoing the simulation from
this starting point, using the same random numbers,
$u_0,\ldots,u_{N-1}$, as before.  If the chain started
from $y_0$ coalesces with the original chain before time $N$, there is
no need to perform any further computations, since each $y_t$ from
that time on will be the same as the corresponding $x_t$.  Here is the
basic procedure (without the diagnostics that will be added in
Section~\ref{sec-diag}), which is also illustrated in
Figure~\ref{fig-basic}:
\begin{list}{}%
{\setlength{\leftmargin}{0.65in}%
\setlength{\labelwidth}{0.25in}%
\setlength{\topsep}{4pt}}
\item[\textbf{Basic circularly-coupled Markov chain simulation:}\hfill]
\item[1)\hfill] Randomly draw $x_0$ from the initial state distribution, $p_0$.
\item[2)\hfill] For $t = 1,\ldots,N$: \\[4pt]
                \hspace*{0.2in}Randomly draw $u_{t-1}$ from the 
                distribution $U$, independently of previous draws. \\
                \hspace*{0.2in}Let $x_t = \phi(x_{t-1},u_{t-1})$.
\item[3)\hfill] Let $y_0=x_N$.
\item[4)\hfill] For $t = 1,\ldots,N$ while $y_{t-1} \ne x_{t-1}$: \\[4pt]
                \hspace*{0.2in}Let $y_t = \phi(y_{t-1},u_{t-1})$.
\item[5)\hfill] Let the remaining $y_t$ be the same as the corresponding $x_t$.
\end{list}

In practice, pseudo-random numbers would generally be used,
eliminating the need to save $u_0,\ldots,u_{N-1}$; instead, the
pseudo-random number generator can simply be re-initialized using the
original seed.  If the amount of memory needed to save a state is
large, the values of $x_0,\ldots,x_{N-1}$ might not be saved when they
are first generated, but instead be recomputed as $y_0,y_1,\ldots$ are
generated.  The time devoted to this recomputation will be relatively
small if coalescence of $y_0,y_1,\ldots$ with $x_0,x_1,\ldots$ is
rapid.  Estimates for the expectations of functions of state that are
of interest can be found without saving all the states using sums of
the values of these functions at all iterations.  These sums can be
accumulated during the initial simulation, and then updated as each
wrapped-around state is simulated.

If the wrapped-around chain fails to coalesce with the original chain
(ie, if $y_N \ne x_N$), the procedure may be seen as having failed.
The project of sampling from $\pi$ might then be abandoned, or the
procedure might be redone with a substantially larger value for $N$.
For purposes of theoretical analysis, however, I will assume that the
$y_t$ for $t=0,\ldots,N\!-\!1$ are always treated as a sample from the
equilibrium distribution of the chain, and used to estimate
expectations of functions with respect to this distribution, even if
coalescence did not occur.  (Note that since $y_N=y_0$ when
coalescence does occur, these points should not both be included in
the sample, as this point would then count double.)  In the next
section, I show that all these $y_t$ will indeed have approximately
the equilibrium distribution, provided certain assumptions regarding
speed of coalescence are satisfied.  Of course, the $y_t$ will
generally be dependent, and this will need to be accounted for when
assessing the accuracy of the estimates obtained, as with standard
Markov chain Monte Carlo procedures.

\subsection{Proof of approximate correctness}\label{sec-proof}\vspace*{-6pt}

The following theorem guarantees the approximate correctness of circular 
coupling, under certain assumptions:

\noindent\textbf{Theorem:}\ \ {\em Let $\pi$, $U$, and $\phi$
be such that $\phi(x,u)\sim\pi$ if $x$ and $u$ are independent
with $x\sim\pi$ and $u\sim U$.
Then each point $y_t$, 
for $t=0,\ldots,N$, that is generated by the basic circularly-coupled 
Markov chain simulation procedure with a given $N$ (assumed here to be
even) has a distribution that is within 
$2\epsilon+\delta$ of the equilibrium distribution, $\pi$, in total
variation distance,\footnote{Here, the total variation distance between
distributions $\mu$ and $\nu$ is defined to be $\sup_A
|\mu(A)-\nu(A)|$, where the supremum is over all events $A$.  Total
variation distance is sometimes given an alternative definition that
is twice this.} provided $\epsilon$ and $\delta$ are such that the 
following conditions hold regarding coupled chains (ie, chains that are
simulated using the same random number sequence, 
$u_0,u_1,\ldots$):

\vspace{-6pt}

\begin{enumerate}
\item[1)] If two chains are started from states drawn from
          the equilibrium distribution, $\pi$, independently of each other,
          and of the $u_t$, they will coalesce within
          $N/2$ iterations with probability at least $1-\epsilon$.
\item[2)] If a chain is started from a state drawn from $\pi$, independently
          of the $u_t$, and another chain is started from a state drawn
          from the distribution $p_0$, independently of the initial state
          of the other chain and of the $u_t$, then the two chains will
          coalesce within $N$ iterations with probability at least 
          $1-\delta$.\vspace{-6pt}
\end{enumerate}}

The proof looks at another way of generating the sequence $y_0,\ldots,y_N$,
along with $x_0,\ldots,x_N$, by means of the procedure given
below, and illustrated in Figure~\ref{fig-proof}.


\begin{list}{}%
{\setlength{\leftmargin}{0.65in}%
\setlength{\labelwidth}{0.25in}%
\setlength{\topsep}{4pt}}
\item[\textbf{Theoretical circular coupling procedure:}\hfill]
\item[1)\hfill] Randomly draw $x_0$ from the initial state distribution, $p_0$.
\item[2)\hfill] For $t = 1,\ldots,N$: \\[4pt]
                \hspace*{0.2in}Randomly draw $u_{t-1}$ from the 
                distribution $U$, independently of previous draws. \\
                \hspace*{0.2in}Let $x_t = \phi(x_{t-1},u_{t-1})$.
\item[3)\hfill] Randomly draw $v_0$ and $w_{N/2}$ from $\pi$, each independently
                of the other and of the $u_t$.
\item[4)\hfill] For $t = 1,\ldots,N/2$:\ \ Let $v_t = \phi(v_{t-1},u_{t-1})$. \\
                For $t = N/2+1,\ldots,N$:\ \ Let $w_t = \phi(w_{t-1},u_{t-1})$.
\item[5)\hfill] Let $v^*_{N/2}=v^{}_{N/2}$ and $w^*_0 = w^{}_N$.
\item[6)\hfill] For $t = N/2+1,\ldots,N$:\ \ Let $v^*_t = 
                                                   \phi(v^*_{t-1},u_{t-1})$.
                \\
                For $t = 1,\ldots,N/2$:\ \ Let $w^*_t = 
                                                   \phi(w^*_{t-1},u_{t-1})$.
\item[7)\hfill] Let $y_t = w^*_t$ for $t=0,\ldots,N/2-1$ and let $y_t = v^*_t$ 
                for $t = N/2,\ldots,N$.
\end{list}
\begin{figure}[t]
\centerline{\psfig{file=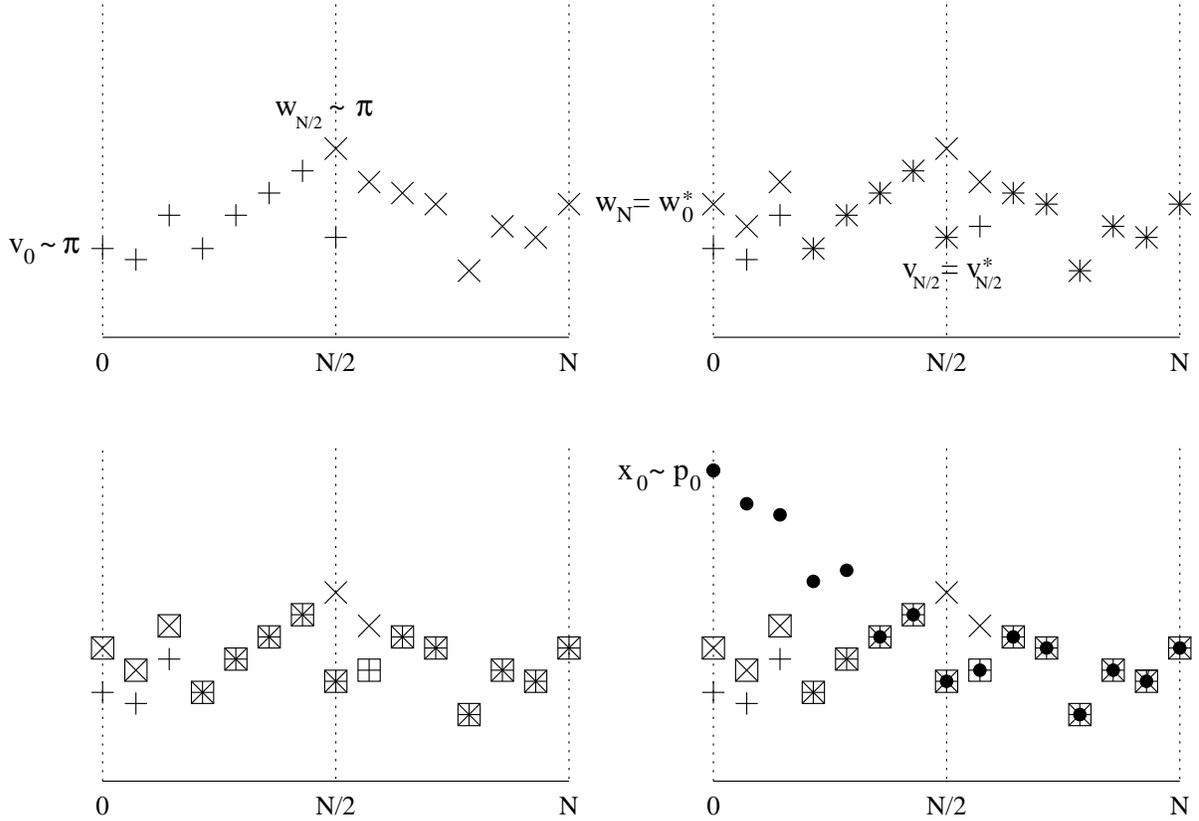,width=6.2in}}
\caption[]{Proof of approximate correctness.  The $v_t$ and $v^*_t$ are shown 
as $+$ signs, the $w_t$ and $w^*_t$ as $\times$ signs, the $y_t$ as squares, 
and the $x_t$ as dots.  The top left shows generation of $v_0,\ldots,v_{N/2}$ 
and $w_{N/2},\ldots,w_N$ starting from states drawn from $\pi$.  
The top right shows the continuation of these
sequences, $v_{N/2}=v^*_{N/2},\ldots,v^*_N$ and $w_N=w^*_0,\ldots,w^*_{N/2}$.
In the bottom left, the sequence $y_0,\ldots,y_N$ used for estimation
is identified.  The bottom right shows a sequence $x_0,\ldots,x_N$ started 
from $p_0$ coalescing with the sequence $y_0,\ldots,y_N$, permitting this
sequence to be found without the need to sample from $\pi$.}\label{fig-proof}
\end{figure}
The generation of $x_0,\ldots,x_N$ in steps (1) and (2) above is the same 
as for the practical circular
coupling procedure of the previous section, but the rest of the theoretical 
procedure could not be carried out in practice, since it requires
sampling directly from $\pi$, which is presumably infeasible.  

However, we can use this theoretical procedure to prove the approximate
correctness of the practical circular coupling procedure.
First, we will see that the $y_t$ produced by the
theoretical procedure all have distribution $\pi$.  Second, when the
two conditions of the theorem hold, we will see that the distribution of 
$y_t$ obtained with the theoretical procedure is approximately the same as 
for the practical procedure.

That each of the $y_t$ obtained using the theoretical procedure has
distribution $\pi$ follows easily as long as $\pi$ is an invariant
distribution of the Markov chain defined by the function $\phi$ and
the random number distribution $U$ --- ie, as long as the distribution
of $\phi(x,u)$ is $\pi$ when $x$ has distribution $\pi$ and $u$ has
distribution $U$, independent of $x$.  From this, it follows that all
the $v_t$ and $v^*_t$ and all $w_t$ and $w^*_t$ have distribution
$\pi$, since they are produced by transitions that leave $\pi$
invariant, starting from a state drawn from $\pi$.  Since each $y_t$
is defined to be equal to either $w^*_t$ or $v^*_t$, the $y_t$ must
all have distribution $\pi$ as well.

For the second part of the proof, we first note that when the two
conditions of the theorem hold, the following events
occur with the indicated probabilities:\vspace{-8pt}
\begin{enumerate}
\item[a)] $v_{N/2}=w^*_{N/2}$ with probability at least 
          $1-\epsilon$.\vspace{-6pt}
\item[b)] $w_N=v^*_N$ with probability at least $1-\epsilon$.\vspace{-6pt}
\item[c)] $x_N=v^*_N$ with probability at least $1-\delta$.\vspace{-6pt}
\end{enumerate}
For event (a), this follows from Condition~(1) because $v^{}_{N/2}$ and 
$w^*_{N/2}$ are the result of $N/2$ iterations with the same $u_t$ 
(for $t=0,\ldots,N/2-1$), starting from points 
$v^{}_0$ and $w^*_0$ that are both drawn independently from $\pi$, independently
of these $u_t$.  That $w^*_0=w^{}_{N}$ is drawn from $\pi$ follows from it
being produced from $w^{}_{N/2}$, which is drawn from $\pi$, by applying 
transitions that leave $\pi$ invariant.  Note that these transitions 
are defined in terms of $u_t$ for $t=N/2,\ldots,N\!-\!1$, which does not 
overlap the set of $u_t$ above.  The bound on the probability of event~(b) 
follows from Condition~(1) similarly, and the bound on the
probability of event~(c) follows from Condition~(2).

The event of (a), (b), and (c) all occurring has probability at least
$1-2\epsilon-\delta$, because the probability that at least one of
(a), (b), and (c) will not occur is at most $2\epsilon+\delta$ (this
bound is valid even though the three events may not be independent).
When this happens, $x^{}_N=v^*_N=w^{}_N=w^*_0=y^{}_0$.  Furthermore,
$y_t=\phi(y_{t-1},u_{t-1})$ for $t=1,\ldots,N$ --- this is obvious for
all $t$ except $N/2$, and for that $t$, it is a consequence of
$v^*_{N/2}=v^{}_{N/2}=w^*_{N/2}$.  When all of (a), (b), and (c)
occur, the values of $y_t$ generated will therefore be the same as
would be generated by the basic circular coupling procedure.

From the coupling inequality (Lindvall 1992), we can bound the total
variation distance between the distribution of $y_0,\ldots,y_{N-1}$ as
produced by the practical procedure and the distribution of
$y_0,\ldots,y_{N-1}$ as produced by the theoretical procedure by the
probability of these sequences differing, which we have seen is in
turn bounded above by $2\epsilon+\delta$.  The total variation
distance between the distributions of any individual $y_t$ for the two
procedures is bounded by the same quantity.  Since the distribution of
each $y_t$ produced by the theoretical procedure is $\pi$, we see that
the distribution of each $y_t$ produced by the practical procedure is
within $2\epsilon+\delta$ of $\pi$ in total variation distance.

\subsection{Testing the conditions for approximate 
correctness}\label{sec-diag}\vspace*{-6pt}

Conditions (1) and (2) required for the approximate correctness of
the circular coupling procedure will seldom be verifiable
theoretically.  Instead, we will have to content ourselves with an
empirical diagnostic test.  

This test starts by tentatively assuming that the two conditions are true
for the value of $N$ we are using, and for some fairly small values of
$\epsilon$ and $\delta$.  If so, the value of each $y_t$ obtained
should come from close to the equilibrium distribution $\pi$.
Although $y_t$ will not be completely independent of
$u_t,u_{t+1},u_{t+2},\ldots$, the dependence should in practice be
sufficiently slight that we can see $y_t,y_{t+1},y_{t+2},\ldots$ as a
realization of the Markov chain started at equilibrium, at least as
long as we look only up to $y_{t+k}$ with $k \ll N$.  (Here and below,
addition and subtraction on times is done modulo $N$, so that if
$t=N\!-\!1$, then $y_{t+1}$ refers to $y_0$.)

Condition~(2) states that such a sequence,
$y_t,y_{t+1},y_{t+2},\ldots$, will with high probability coalesce
within $N$ iterations with another coupled chain started in a state
drawn from the initial state distribution $p_0$.  Choosing some $r$
that divides $N$, we can test whether this in fact occurs when
starting at times $t=N/r,\,2N/r,\,\ldots,\,(r\!-\!1)N/r$ by simulating
auxiliary chains starting at those times.  In practice, we would
usually wish for coalescence to occur in considerably fewer than $N$
iterations, so let us suppose that we simulate each such chain for
only some number $k < N/2$ iterations, or until the auxiliary chain
coalesces with $y_t,y_{t+1},y_{t+2},\ldots$

This leads to the following extension of the basic circular coupling
procedure:
\begin{list}{}%
{\setlength{\leftmargin}{0.65in}%
\setlength{\labelwidth}{0.25in}%
\setlength{\topsep}{4pt}}
\item[\textbf{Circularly-coupled simulation with 
              auxiliary diagnostic chains:\hspace*{-10pt}}\hfill]

\item[1-5)\hfill] Perform steps (1) to (5) of the basic circularly-coupled
                  Markov chain simulation procedure.

\item[6)\hfill] Let $c_0$ be the number of steps needed for 
                the wrapped-around chain to coalesce with the original chain
                --- ie, let $c_0$ be the smallest $t$ such that 
                $y_t=x_t$ --- unless the chains do not coalesce within $k$
                iterations, in which case let $c_0=k$.
                
\item[7)\hfill] For $i=1,\ldots,r\!-\!1$: \\[4pt]
                \hspace*{0.2in}Let $s=iN/r$.\\[4pt]
                \hspace*{0.2in}Randomly draw $z_{i,s}$ from the initial 
                               state distribution, $p_0$. \\[4pt]
                \hspace*{0.2in}Set $c_i = 0$.\\[4pt]
                \hspace*{0.2in}For $t=s\!+\!1,\ldots,s\!+\!k$ (modulo $N$)
                                 while $z_{i,t-1} \ne y_{t-1}$:\\
                \hspace*{0.4in}  Let $z_{i,t} = \phi(z_{i,t-1},u_{t-1})$.\\
                \hspace*{0.4in}  Set $c_i = c_i + 1$.
\end{list}
The time required for this procedure will be roughly proportional to the
number of Markov chain iterations (ie, the number of evaluations of $\phi$),
which will be $N + \sum_i c_i$.
Note that if all the auxiliary chains coalesce with the wrapped-around
chain, $y_0,\ldots,y_N$, there will be no real distinction
between the auxiliary chains and the original chain that was started 
with $x_0$ drawn from $p_0$.  The same wrapped-around chain
would have been found from any of the $r$ starting points.

The values of $c_i$ for $i=0,\ldots,r\!-\!1$ that are obtained by this
procedure are indicative of whether Condition~(2) holds.  If many of
the $c_i$ are equal to $k$, we should not be confident that this
condition holds, and should rerun the simulation with a larger value
for $k$ and probably a larger value for $N$ as well (recall that $k$
should be substantially smaller than $N$).  It may often be reasonable
to think that the $c_i$ have an approximately geometric distribution,
in which case the parameter of this distribution could be estimated
from this (right-censored) data, and used to estimate the value of
$\delta$ for which Condition~(2) holds.

This test does not provide direct information about Condition~(1),
which involves two chains started from the equilibrium distribution.
However, if the evidence from the auxiliary chains leads us to
conclude that all but a small fraction, $q$, of chains started from
$p_0$ coalesce in no more than $N/2$ iterations with a chain started
from the equilibrium distribution, then we can also conclude that two
chains started from the equilibrium distribution will coalesce with
each other within $N/2$ iterations with probability at least $1-2q$,
since if they both coalesce with a chain started from $p_0$, they must
also coalesce with each other.

If all the auxiliary diagnostic chains are observed to coalesce
quickly with the wrapped-around chain, we therefore have reason to
believe that both conditions for approximate correctness hold.  This
will not be an absolute guarantee, however.  It could be that the
initial state distribution, $p_0$, gives little probability to a
region that has high probability under $\pi$, and that is isolated
from the regions that do have high probability under $p_0$.  Both the
wrapped-around chain and the auxiliary diagnostic chains might never
visit this isolated region, in which case the diagnostic chains would
present a self-consistent but drastically incorrect picture of the
distribution of coalescence times.  To help avoid this, it is
desirable for $p_0$ to be ``overdispersed'' with respect to $\pi$, but
even if this is so, there is no guarantee that all high probability
regions of $\pi$ will be found, since some region with a large
probability under $\pi$ might have a small ``basin of attraction'',
and hence could be missed even if it is within the high probability
region of $p_0$.  This can occur, for example, when $\pi$ gives
substantial probability to a small region with very high probability
density, which the chain is unlikely to chance upon.

\subsection{Parallel simulation}\label{sec-par}\vspace*{-6pt}

The auxiliary chains used as diagnostics in the previous section are
expected to coalesce with the wrapped-around chain reasonably rapidly.
If this is indeed so, coalescence of each auxiliary chain with the
next auxiliary chain (started $N/r$ time steps forward) will also be
fairly rapid.  It will then be possible to find the wrapped-around
chain by parallel computation on several processors, in less time than
would be needed to simulate the wrapped-around chain using a single
processor.  The following procedure is based on this idea:


\begin{list}{}%
{\setlength{\leftmargin}{0.65in}%
\setlength{\labelwidth}{0.25in}%
\setlength{\topsep}{4pt}}
\item[\textbf{Parallel simulation of a circularly-coupled Markov 
      chain:\hspace*{-10pt}}\hfill]
\item[In parallel, processors numbered by $i = 0,\ldots,r\!-\!1$ do
      the following:\hfill]~\vspace*{-8pt}

\item[(The variables $s$, $t$, and $z$ are local to each processor)\hfill]
\item[1)] Let $s=iN/r$.
\item[2)] For $t=s,\,\ldots,\,s+N/r-1$:\\[4pt]
          \hspace*{0.2in}Randomly draw $u_t$ from the distribution $U$, 
          independently of other draws.
\item[3)] Randomly draw $y_s$ from the distribution $p_0$, independently
          of other draws.
\item[4)] For $t=s+1,\,\ldots,\,s+N/r-1$:\\[4pt]
\hspace*{0.2in}Set $y_t = \phi(y_{t-1},u_{t-1})$.
\item[5)] Set $z=\phi(y_{s+N/r-1},u_{s+N/r-1})$.
\item[6)] Send $z$ to
            processor $i\!+\!1$ (modulo $r$) as the new value for $y_{s+N/r}$ .
\item[7)] Repeat the following: \\[4pt]
\hspace*{0.2in}Wait for a new value for $y_s$ to be received from processor 
               $i\!-\!1$ (modulo $r$).\\[4pt]
\hspace*{0.2in}For $t=s+1,\,\ldots,\,s+N/r-1$ 
                 while $y_t \ne \phi(y_{t-1},u_{t-1})$:\\[4pt]
\hspace*{0.4in}Set $y_t = \phi(y_{t-1},u_{t-1})$.\\[4pt]
\hspace*{0.2in}If $z \ne \phi(y_{s+N/r-1},u_{s+N/r-1})$:\\[4pt]
\hspace*{0.4in}Set $z=\phi(y_{s+N/r-1},u_{s+N/r-1})$.\\
\hspace*{0.4in}Send $z$ to
          processor $i+1$ (modulo $r$) as the new value for $y_{s+N/r}$ .
\item[The procedure terminates when all processors are waiting.\hfill]
\end{list}
The wrapped-around chain consists of the final values of
$y_0,\ldots,y_{N-1}$ that are stored in the $r$ processors.  
The subscript of $y$ wraps around in the above procedure, so that
processor $r\!-\!1$ sends $y_N$ to processor $0$, which uses it to replace
the old value of $y_0$.  An ordinary, non-circular Markov chain 
simulation can be performed in parallel in the same way as above by omitting 
this wrap-around, keeping $y_0$ fixed at its original value, but I will
not discuss this further here.

The computation time for the above simulation will be at least the
time required to simulate $N/r$ Markov chain iterations, since that
many iterations will always be done in steps~(4) and (5).  Each
processor will then begin simulating a chain starting from the new
value received for its $y_s$.  If each of these chains coalesces
within $N/r$ iterations with the chains that were simulated starting
from the original values for each $y_s$, then each processor will find
that the value for $y_{s+N/r}$ that it communicates to the next
processor is unchanged, and the entire procedure will terminate.  The
time taken will be that required for between $N/r$ and $2N/r$ Markov
chain iterations.

If, on the contrary, not all of these chains coalesce within $N/r$
iterations, one or more of the processors will have to perform a third
simulation.  In general, a processor might have to rerun its
simulation any number of times, as a result of the previous processor
sending it new start states.  Assuming that a sequential simulation
would have resulted in the wrapped-around chain coalescing with the
original chain, the time required for the parallel simulation will be
roughly proportional to the maximum number of iterations that any of
the $r$ chains with different starting points take to coalesce with
this wrapped-around chain.  If this time is comparable to the time for
a sequential simulation, the slow coalescence of the auxiliary chain
would be indicative of a problem, and it would usually be best to stop
the whole procedure, and restart it with a larger value for $N$.

It is possible that the procedure as shown will not terminate.  This
will happen if different starting points lead to different
wrapped-around chains; Figure~\ref{fig-ex4} in Section~\ref{sec-demo}
below illustrates this possibility.  In practice, the procedure should
be terminated when some processor has received more than some maximum
of new values for its starting point.  The maximum number of such new
starting points is a rough diagnostic of how rapidly the chains
couple, providing information similar to that provided by the $c_i$ in
the procedure of Section~\ref{sec-diag}.

This parallel simulation procedure may be adaptable to vector
computation, provided the computation of $\phi$ does not involve lots
of conditional computations.  Such a vectorized simulation might be
appropriate when vector operations are supported by hardware, or when
programming is done in an interpreted language in which vector
operations are not much slower than scalar operations, due to the
fixed overhead of interpretation.

Finally, note that although the parallel simulation procedure aims to
do roughly what is done by the sequential procedure with auxiliary
diagnostic chains of Section~\ref{sec-diag}, the actual computations
done may differ.  When not all chains coalesce with the wrapped-around
chain within $N/r$ iterations, the sequential procedure of
Section~\ref{sec-diag} will simulate two or more auxiliary chains that
operate at the same times.  It is possible that these chains will
coalesce with each other before coalescing with the wrapped-around
chain, but the sequential procedure does not detect this, and will
simulate the coalesced chains separately.  Coalescence of chains is
detected differently in the parallel procedure, however, which may
lead to such portions of chains being simulated only once.  Use of
this procedure may therefore be advantageous even when one has
available only a single processor (which executes the $r$ parallel
processes by time sharing).

\section{Coupling using random-grid Metropolis
updates}\label{sec-rg}\vspace*{-10pt}

To be practically useful, the circular coupling procedure needs a
representation of the Markov chain transitions in terms of a $\phi$
function that is easily computable and that leads to rapid coalescence
of chains.  Exact sampling methods such as coupling from the past
(Propp and Wilson 1996) also require such a representation, as does
Johnson's (1996, 1998) convergence diagnostic.  Several such schemes
have recently been discussed in the context of exact sampling by Green
and Murdoch (1998).  Note that we are at liberty to choose the Markov
chain transitions to facilitate coupling, though this might sometimes
come at a cost in terms of convergence rate.

In this section, I present a simple coupling scheme based on
``random-grid'' Metropolis updates, which can produce exact
coalescence in continuous state spaces, and use it to demonstrate some
aspects of circular coupling using simple one-dimensional
distributions.  I also show that the random-grid method alone does not
work well for high-dimensional problems.  However, random-grid
Metropolis updates can be combined with other standard methods to
obtain better results, as is discussed in Section~\ref{sec-schemes},
and applied to the example in Section~\ref{sec-lr}.

\subsection{Random-grid Metropolis in one
dimension}\label{sec-rg1}\vspace*{-6pt}

Suppose that our state, $x$, consists of a single real value, and that
our desired distribution is given by the density function $\pi(x)$.
Recall that the Metropolis algorithm defines a Markov chain transition
in terms of a density function, $g(x^*|x)$, for proposing a move to
state $x^*$, given that the chain is currently in state $x$.  This
proposal is accepted with probability $\min[1,\,\pi(x^*)/\pi(x)]$.  If
the proposal is rejected, the new state is the same as the old state.
Provided that the proposal distribution is symmetrical (ie,
$g(x^*|x)=g(x|x^*)$), this update leaves the distribution $\pi$
invariant.  

I will consider a random-walk Metropolis algorithm using proposals that 
are uniformly distributed in an
interval of width $2w$, centred on the current state, for which
\beq 
  g(x^*|x) & = & \left\{\begin{array}{ll}1/(2w) & \mbox{if $|x^*-x|<w$} \\
                                         0   & \mbox{otherwise}
                 \end{array}\right.
\label{eq-prop}\eeq
The most obvious way of expressing this update in terms of a function
$\phi$ is as follows:
\beq
    \phi(x,u) & = & \left\{\begin{array}{ll}  
       x+2w\,(u_1-1/2) & \mbox{if $u_0 < \pi(x+2w\,(u_1-1/2))\,/\,\pi(x)$} \\
       x            & \mbox{otherwise}
    \end{array}\right.
\label{eq-couple0}\eeq
This function takes the current state and a vector, $u=(u_0,u_1)$, of two
$\mbox{Uniform}(0,1)$ random numbers as arguments, and returns the
next state, which will have the distribution as defined for the
Metropolis algorithm with the proposal distribution above.  However, with
this $\phi$ function, the probability of exact coalescence of chains is zero,
whenever the current states are distinct.

Fortunately, the same Markov chain transition probabilities are
obtained with the following $\phi$ function, which can lead to
coalescence with positive probability:
\beq
    \phi(x,u) & = & \left\{\begin{array}{ll}  
       f(x,u) & \mbox{if $u_0 < \pi(f(x,u))\,/\,\pi(x)$} 
       \\
       x            & \mbox{otherwise}
    \end{array}\right.
\label{eq-couple}\\[6pt]
 \lefteqn{\mbox{where 
  $f(x,u)\ =\ 2w\,[(u_1-1/2)+\mbox{Round}(x/(2w)-(u_1-1/2))]$}}\ \ \ \ \ \ 
\nonumber
\eeq
Round returns the integer nearest its argument.  The function $f(x,u)$
can be seen as first transforming the state to $x'=x/(2w)-(u_1-1/2)$,
then rounding to the nearest integer, and finally applying the inverse
transformation.  For a given value of $u_1$, a range of
values for $x$ of width $2w$ all result in the same value for
$f(x,u)$.  Hence, two chains whose current states are in this range, and 
which both accept the proposed point, will coalesce exactly.

This $\phi$ function can be visualized as first laying down a grid of
points spaced $2w$ apart, with the position of the grid being chosen
uniformly at random, and then proposing to move to the point on this
grid that is nearest the current state, $x$.  It is then clear that
the distribution of the proposed state is as in
equation~(\ref{eq-prop}).  

How well can this method be expected to work?  There are two aspects
to this question --- how rapidly a Markov chain using these updates
will converge to the equilibrium distribution, and how rapidly two
chains coupled in this way will coalesce.  As for any coupling method,
the average coalescence time cannot be less than the time for
approximate convergence, but it might well be larger, if the coupling
scheme is ill-chosen.

The convergence rate for random-grid Metropolis will of course depend
on the distribution, and on the choice of $w$.  For one-dimensional
problems, the optimal $w$ is generally quite large, even if this leads
to a low acceptance rate.  However, I will here assume that $w$ is
chosen so as to produce a fairly high acceptance rate, since this is
more relevant to the higher-dimensional problems examined later.

If the acceptance rate is substantially greater than one-half (eg,
3/4), coalescence should occur about as rapidly as convergence to the
equilibrium distribution.  With such a high acceptance rate, two
coupled chains must often accept simultaneously, which will have a
good chance of causing coalescence if the states of the two chains are
substantially less than $2w$ apart.  Furthermore, the chains will
indeed approach to within that distance fairly often, because they
both move in steps less than $w$ in size, and in one dimension, their
paths cannot avoid crossing.  (One chain cannot stay below the other
indefinitely, since both chains must be sampling from the same
equilibrium distribution.)

Before discussing how this scheme can be generalized to higher
dimensions, I first use one-dimensional random-grid updates
to demonstrate some general aspects of circular coupling.

\subsection{Simple demonstrations using random grid
Metropolis}\label{sec-demo}\vspace*{-6pt}

\begin{figure}[t]

\centerline{\psfig{file=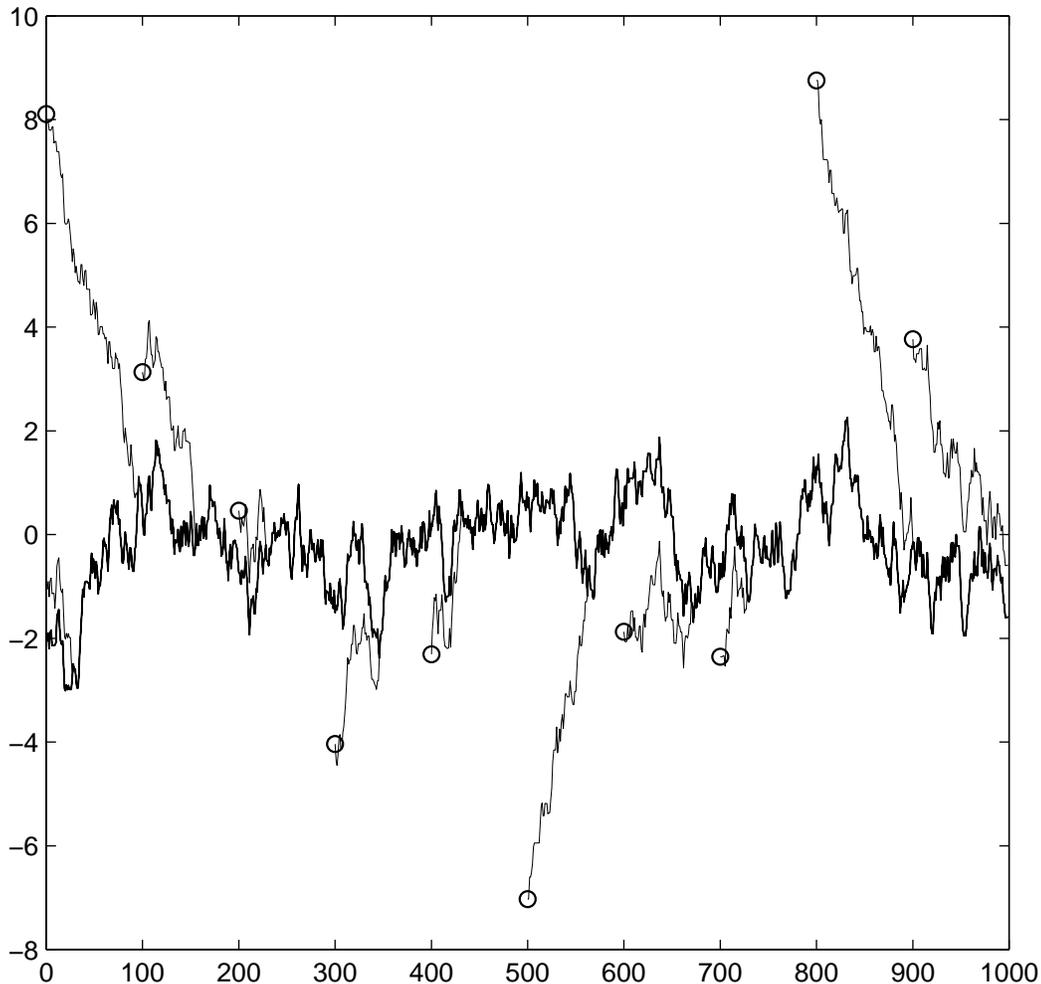}}

\caption[]{A circularly-coupled random-walk Metropolis simulation sampling 
from the $N(0,1)$ distribution.  Ten chains started with states drawn from
the initial distribution $N(0,5^2)$ (shown as circles) all coalesce with the 
wrapped-around chain (the thick line) in less than 150
iterations, much less than the total of 1000.  This is consistent with the 
conditions required for the circular coupling procedure to be approximately 
correct.}\label{fig-ex1}

\end{figure}

\begin{figure}[t]

\centerline{\psfig{file=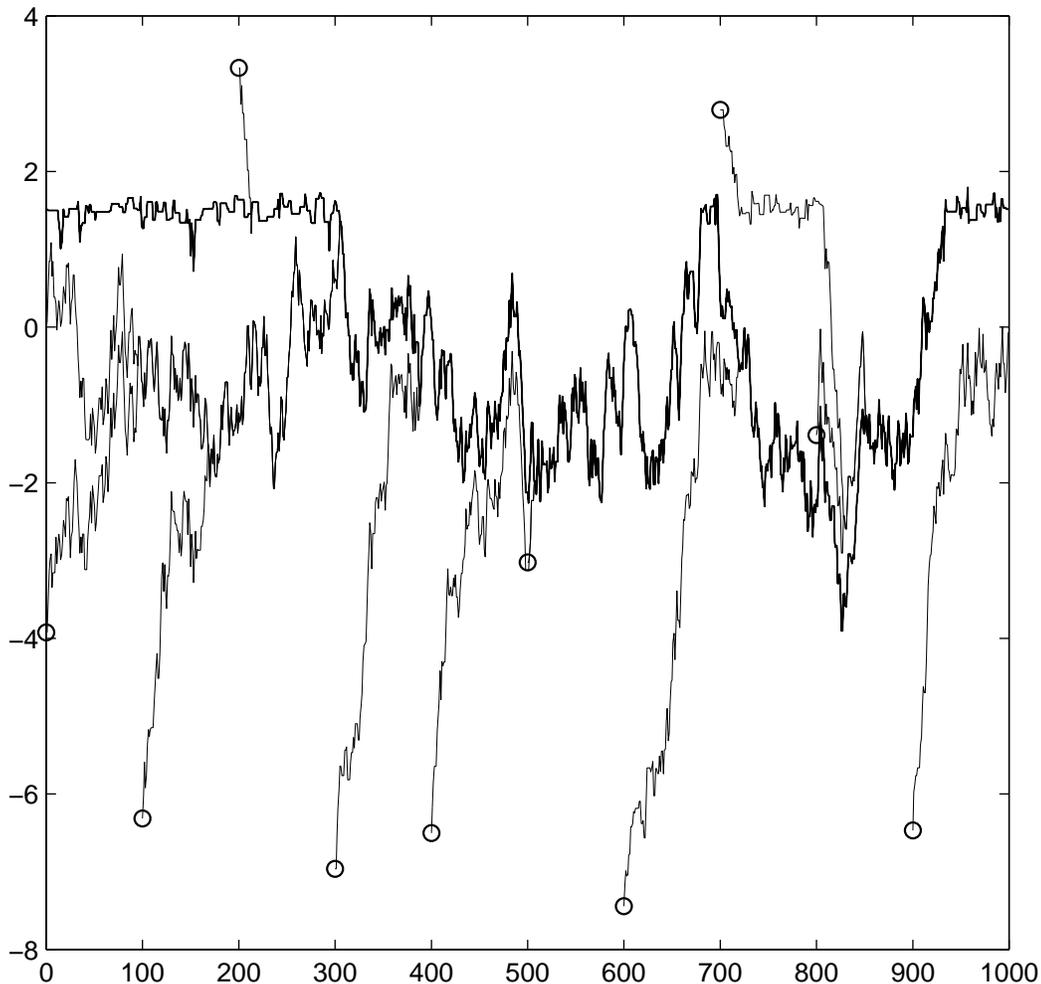}}

\caption[]{A circularly-coupled random-walk Metropolis simulation sampling 
from the distribution $(3/4) N(-1,1) + (1/4) N(1.5,0.1^2)$.  One of the ten
chains started from the initial distribution $N(0,5^2)$ takes 400 iterations
to coalesce with the wrapped-around chain.  Since this is a substantial
fraction of the total of 1000 iterations, one might doubt whether the
conditions for the circular coupling procedure to be approximately correct
are satisfied.}\label{fig-ex2}

\end{figure}

\begin{figure}[t]

\centerline{\psfig{file=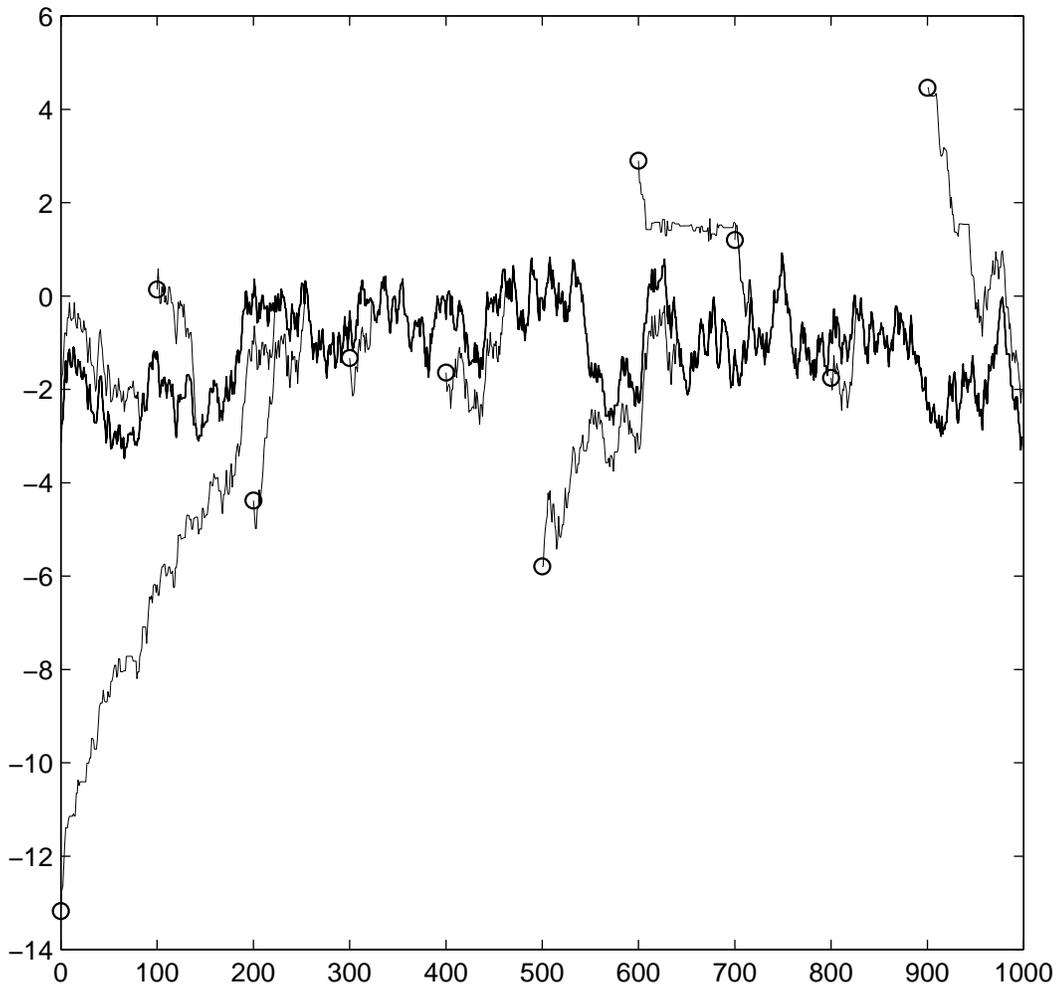}}

\caption[]{Another run of the circularly-coupled simulation shown in
Figure~\ref{fig-ex2}.  This time, all ten chains  coalesce with a
wrapped-around chain that visits only the lower of the two 
modes.}\label{fig-ex3}

\end{figure}

\begin{figure}[t]

\centerline{\psfig{file=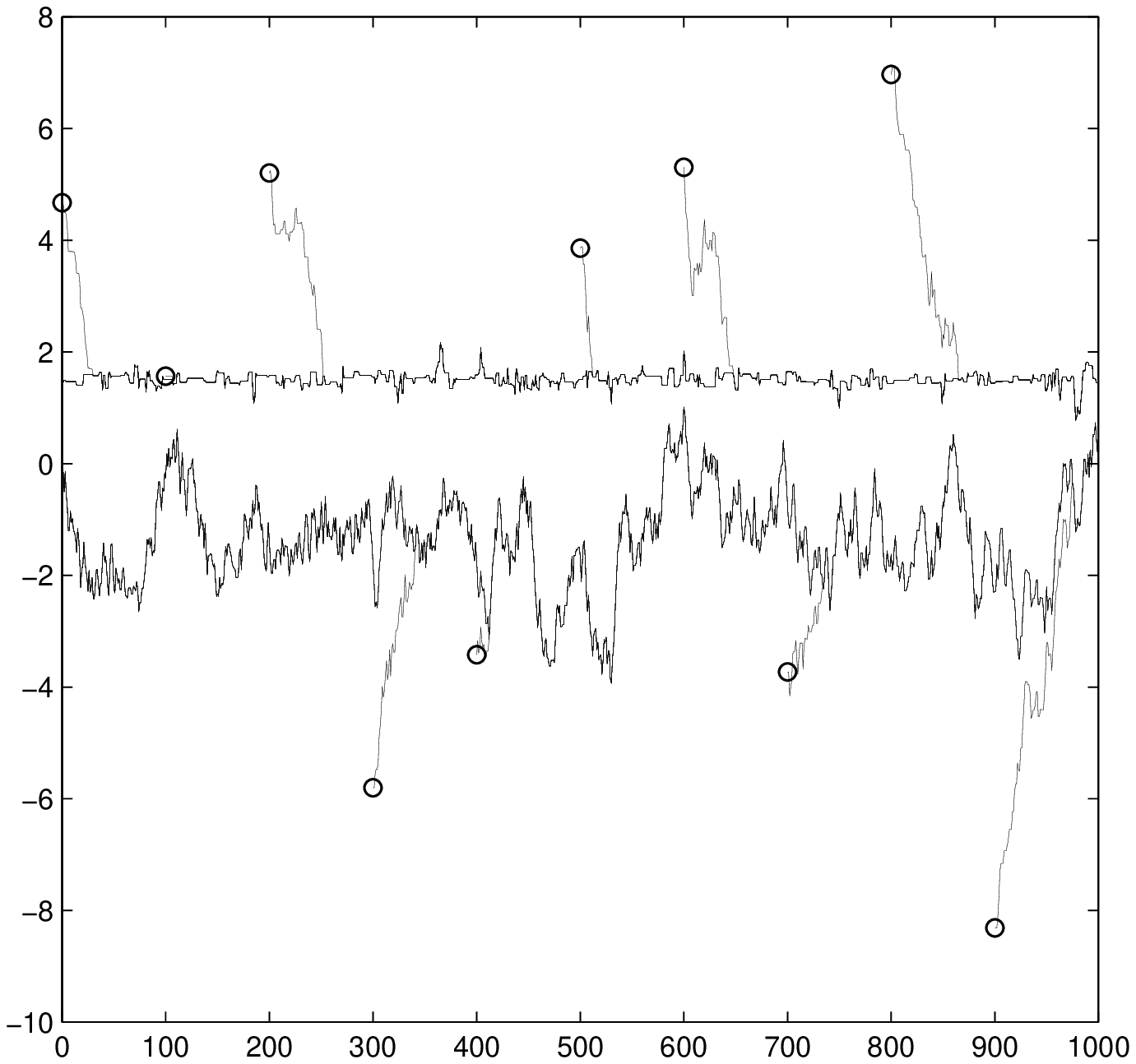}}

\caption[]{A run of the same circularly-coupled simulation as in
Figure~\ref{fig-ex2} in which a single wrapped-around chain was not found
after simulating each chain for $N$ iterations.  Instead, six of the
chains coalesced into a wrapped-around chain that stays in the
upper mode, while the other four chains coalesced into a wrapped-around 
chain that stays in the lower mode.}\label{fig-ex4}

\end{figure}

To illustrate the concept of circularly-coupled Markov chain
simulation, I include here two simple illustrations with a
one-dimensional state space.  Both examples use the random-grid
Metropolis algorithm of the previous section, which is based on the
uniform proposal distribution of equation~(\ref{eq-prop}), with
$w=1/2$, coupled using the scheme of equation~(\ref{eq-couple}).
This value for $w$ was chosen deliberately so as to produce a
somewhat inefficient sampler.

The first example illustrates the behaviour of circular coupling when
coalescence is rapid.  Figure~\ref{fig-ex1} shows a simulation of
length $N=1000$ that samples from the $N(0,1)$ distribution, with
$N(0,5^2)$ as the initial state distribution.  Ten chains were
simulated in total --- one started at $t=0$, plus nine auxiliary
chains started at $t=100,200,\ldots,900$.  All chains coalesce rapidly
with the wrapped-around chain, an indication (but not an absolute
guarantee) that the conditions for the states of the wrapped-around
chain to all come from the equilibrium distribution are satisfied.

The second example illustrates how circular coupling behaves when coalescence
is less rapid.  The distribution to be sampled from is in this case a bimodal
mixture of normals, $(3/4) N(-1,1) + (1/4) N(1.5,0.1^2)$.  Figure~\ref{fig-ex2}
shows a circularly-coupled simulation for this distribution that is typical
of runs of length $N=1000$, with $r=10$ starting points.
Some of the ten chains started with states from the $N(0,5^2)$ distribution
coalesce rapidly, as for the first example, but others do not.  In particular,
the chain started at $t=900$ takes about 400 iterations to coalesce with the
wrapped-around chain.  This is evidence that the conditions needed for the
states of the wrapped-around chain to come from approximately the correct
distribution may not be satisfied.

Figure~\ref{fig-ex3} shows another run of the same simulation, with a
different pseudo-random number seed.  This run might well be
misleading, since the wrapped-around chain found samples from only the
lower of the mixture distribution's two modes.  Moreover, the chains
from all ten starting points coalesce with this wrapped-around chain
reasonably quickly, which might lead one to think that the procedure
is sampling from approximately the correct distribution.  The
distribution of states produced by this procedure might indeed be at
least roughly correct, since as seen in the previous figure, other
runs of the same simulation procedure do produce wrapped-around chains
that visit the upper mode as well.  However, even if the distribution
of each state of the wrapped-around chain for $N=1000$ is close to correct,
the states in this chain are clearly highly dependent, and hence a
single wrapped-around chain can fail to provide an adequate sample.
Behaviour similar to this occurs in a few percent of the runs.  Using
more than ten random starting points would reduce the chances of such
a problem remaining undiagnosed.

Figure~\ref{fig-ex4} shows another possible result of the simulation
procedure, which occurs about once in a thousand runs.  Here, six of the ten
chains coalesce to a wrapped-around chain that moves within the upper
mode only.  The chains started from the other four initial states
coalesce to a wrapped-around chain that moves within the lower mode.
When this situation occurs, one can tell that the simulation should
be rerun with a larger value of $N$.

\subsection{Random-grid Metropolis in more than one
dimension}\label{sec-rgmany}\vspace*{-6pt}

The random-grid Metropolis procedure can be used to sample from
multi-dimensional distributions in two ways.  One way is to update one
component of the state at a time, in sequence or randomly, using the
one-dimensional random-grid method of Section~\ref{sec-rg1}.  The
other way is to update all components of the state at once, using a
multi-dimensional random grid.

Multi-dimensional random-grid Metropolis updates use a
proposal distribution that is uniform over a hypercube
centred on the current state, with sides of length $2w$.  To produce
exact coalescence, this is implemented using a $\phi$ function
analogous to that of equation~(\ref{eq-couple}), which can be
visualized as randomly positioning a multi-dimensional square grid
with points separated by $2w$ in each coordinate direction, and then
proposing to move to the grid point nearest the current state.  In
detail, a $d$-dimensional update from state $x$ is done by generating
a vector of $d+1$ independent Uniform(0,1) random variates,
$u=(u_0,u_1,\ldots,u_d)$, and then setting the next state to 
\beq
    \phi(x,u) & = & \left\{\begin{array}{ll}  
       f(x,u) & \mbox{if $u_0 < \pi(f(x,u))\,/\,\pi(x)$} 
       \\
       x            & \mbox{otherwise}
    \end{array}\right.
\eeq
where the function $f(x,u)$ is defined by
\beq
  [f(x,u)]_i & = & 2w\,[(u_i-1/2)+\mbox{Round}(x_i/(2w)-(u_i-1/2))],\ \ \
  \mbox{for $i=1,\ldots,d$}\ 
\eeq
(Note that subscripts of $x$ will now index components of the
multivariate state, not time, as was the case before.)

In contrast to the situation in one dimension, it is conceivable that
two coupled chains with a higher-dimensional state might both sample
from the same equilibrium distribution, moving in steps of size less
than $w$ in each coordinate direction, and yet almost always be
separated by a distance greater than $2w$ in at least one direction,
thereby making exact coalescence unlikely when using the random-grid
procedure.  This will be avoided if coupled chains have a tendency to
approach each other more and more closely, even when they are far
apart, so that they will eventually get close enough for exact
coalescence to occur.

It turns out that both single-component and multi-dimensional
random-grid Metropolis methods do have such a tendency for distant
chains to move closer together.  Unfortunately, this tendency is lost
once the chains are fairly close, so exact coalescence can be quite
delayed.  In the next two sections, I analyse this situation for the
case of sampling from a multivariate normal distribution, and
demonstrate how well the methods work for a nine-dimensional example.
These results provide insight into how to do better by combining
random-grid Metropolis with other updates, as discussed in
Section~\ref{sec-schemes}, but impatient readers may skip to
Section~\ref{sec-schemes} immediately without serious loss of
continuity.

\subsection{Analysis of random-grid Metropolis for multivariate 
normals}\label{sec-rgmvna}\vspace*{-6pt}

It is possible to see that random-grid Metropolis will indeed tend to
bring two coupled chains closer together in the simple case of
sampling from a multivariate normal distribution.  In analysing this
situation, I will assume that times when one of the two chains accepts
its proposal and the other rejects are fairly rare, and that
consequently, when this does occur, it is usually from states that
were produced by both chains accepting a proposal simultaneously.  If
components are updated one at a time, the assumption is that updates
in which one chain accepts a change to a component and the other
rejects are from a state in which the last change to that particular
component was an update in which both chains accepted.  Note that this
assumption will certainly hold if the acceptance rate is quite high,
as it will be if $w$ is chosen to be small.

Under this assumption, we need consider only the following
four possibilities for how the two chains are updated:
\begin{enumerate}
\item Both chains accept their proposals, starting from states
      that are the result of both chains accepting the previous 
      proposal (for the whole state, or for the component currently
      being updated).
\item Both chains reject their proposals.
\item One chain accepts its proposal, but the other rejects,
      starting from states that are the result of both chains accepting. 
\item Both chains accept their proposals, starting from states
      resulting from one chain accepting and the other rejecting.
\end{enumerate}

Let the states of the two chains be $x$ and $x'$, with components
$x_i$ and $x'_i$, for $i=1,\ldots,n$.  Let the vector of differences
in components be $d=|x'-x|$.  After an update for component $i$ in
which both chains accepted, $d_i$, will be a multiple of $2w$, since
the points accepted by the chains were on a grid with this spacing.
Because of this, when this component is next updated, the amount by
which it is altered in the proposal will be the \textit{same} for both
chains, because the offsets of the nearest grid point from the current
states will be identical when the current states are themselves on a
grid with the same spacing.  If both chains accept
this proposal, $d_i$ will be unchanged.  Hence updates of type (1)
above do not change the distance between the chains.

Updates of type (2) obviously do not change the distance between chains.

Updates of type (4) can change the distance between chains, either
increasing or decreasing $d_i$, but the expected value of $d_i$ after
such an update, averaging over the random grid placement, is the same
as the previous value of $d_i$.  To see this, write $d_i = 2wk+h$,
with $k$ an integer and $h \in [0,2w)$.  One can easily see that if
both chains accept, then with probability $h/2w$, the distance between
the new states will be $2w(k+1)$ and with probability $1-h/2w$ this
distance will be $2wk$, giving an expected distance of $(h/2w)2w(k+1)
+(1-h/2w)2wk = 2wk+h = d_i$.  These updates will therefore have no
systematic tendency to move the chains together or apart, even though
the eventual exact coalescence of the chains will occur with a type
(4) update.

It is the effect of updates of type (3) that is crucial.  Since these
updates start with states for which each $d_i$ is a multiple of $2w$,
as for type (1) updates, the changes proposed for each component, $\delta_i$, 
will be the same for both chains.  (Note that if one component is being
updated at a time, $\delta_i$ will be zero for all except the component
currently being updated.)  The acceptance probability for a
proposal to change $x$ by $\delta$ can be written as \beq
 a(x,\delta) 
  & = & \min[1,\,\pi(x+\delta)/\pi(x)] 
  \ \ =\ \ \min[1,\,\exp(-(E(x+\delta)-E(x)))]
\label{eq-axdelta}\eeq
Here, $E(x)=x^{T}\Sigma^{-1}x/2$, where $\Sigma$ is the covariance
of the multivariate normal distribution being sampled (the mean 
is assumed, without loss of generality, to be zero).  We have that
\beq
 E(x+\delta)-E(x) & = & x^{T}\Sigma^{-1}\delta\ +\ \delta^{T}\Sigma^{-1}\delta/2
\eeq
For a chain with state $x$ to accept the update while a chain with state
$x'$ rejects, it must be that $E(x+\delta)-E(x)<E(x'+\delta)-E(x')$,
from which it follows that
\beq
 (x-x')^{T}\Sigma^{-1}\delta & < & 0
\eeq
The quantity on the left can be interpreted as an inner product, using
the metric defined by the positive-definite matrix $\Sigma^{-1}$.
That this inner product is negative indicates that an infinitesimal
change to $x$ in the direction of $\delta$ would reduce the squared
distance between $x$ and $x'$, as measured by
$(x-x')^{T}\Sigma^{-1}(x-x')$.  When the chain with state $x$ accepts
(while the other rejects), its state changes by the finite amount
$\delta$, and though this change is in the right direction, it may
overshoot.  The actual change in the squared distance will be
\beq
  2(x-x')^{T}\Sigma^{-1}\delta\ +\ \delta^{T}\Sigma^{-1}\delta
\label{eq-change}
\eeq
When the chains are far apart, and hence $x-x'$ is large, the first
term in (\ref{eq-change}), which is negative, will likely dominate,
but once the chains are close together, exact coalescence will be
delayed by the effect of the second term, which is always positive.

The rate at which distant chains approach each other can be found by
considering the probability of one chain accepting and the other
rejecting.  If the chains' states are $x$ and $x'$, this probability
is $|a(x,\delta)-a(x',\delta)|$.  (Recall that the acceptance
decisions are linked by use of the same value for $u_0$.)  Changes by
$\delta$ and $-\delta$ are equally likely to be proposed, so we can
average over these two possibilities, while keeping $\delta$ fixed
apart from this overall sign flip. Suppose that $x-x'$ is large
compared to $\delta$, and also that the acceptance rate is high, so
that the exponential in equation~(\ref{eq-axdelta}) can be
approximated linearly.  We can then consider the three cases of such
proposals leading to an increase in $E$ for both chains, a decrease
for both chains, or an increase for one and a decrease for the other,
obtaining \beq
  E_{\pm\delta} \
  |a(x,\delta)-a(x',\delta)| & \approx & |(x-x')^{T}\Sigma^{-1}\delta|/2
\eeq
Multiplying this by the change in squared distance when one chain does
accept while the other rejects (from~(\ref{eq-change}), ignoring the second
term), we find that the expected decrease in squared distance (in the 
metric defined by $\Sigma^{-1}$) is approximately
\beq
  \left[\, (x-x')^{T}\Sigma^{-1}\delta \,\right]^2 & = &
  (x-x')^{T}\Sigma^{-1}\,[\,\delta\delta^{T}\,]\,\Sigma^{-1}(x-x')
\eeq
The expectation of $\delta\delta^{T}$ is $\omega I$, where $\omega=w^2/3$
for multi-dimensional random-grid updates, and $\omega=w^2/3n$ for a random-grid
update of a single component chosen randomly.  Achieving a given acceptance
rate generally requires a value of $w$ for multi-dimensional updates that
is a factor of $\sqrt{n}$ smaller than for single-component updates (Roberts,
\textit{et al.} 1997), so $\omega$ can be considered to be the same when 
comparing the two types of updates.  The expected decrease in
squared distance averaging over the distribution of $\delta$ can now
be found as
\beq
  (x-x')^{T}\Sigma^{-1}\,[\,\omega I\,]\,\Sigma^{-1}(x-x')
  & = & \omega (x-x')^{T}\Sigma^{-2}(x-x')
\label{eq-edec}\eeq

Let the eigenvectors of $\Sigma^{-1}$ be $v_1,\,v_2,\,\ldots,\,v_n$,
normalized to unit length, and let the corresponding eigenvalues be
$\lambda_1 \ge \lambda_2 \ge \cdots \ge \lambda_n > 0$.  The eigenvectors
of $\Sigma^{-2}$ will be the same, with its eigenvalues being $\lambda_1^2
\ge \lambda_2^2 \ge \cdots \ge \lambda_n^2$.  By expressing $x-x'$ in
this eigenvector basis, as $a_1v_1 + a_2v_2 + \cdots + a_nv_n$, we can
write the expected relative change in the squared distance resulting from 
an update as
\beq
 {\omega (x-x')^{T}\Sigma^{-2}(x-x') \over (x-x')^{T}\Sigma^{-1}(x-x')}
 & = & \omega {a_1^2\lambda_1^2 + a_2^2\lambda_2^2 + \cdots + a_n^2\lambda_n^2
         \over a_1^2\lambda_1 + a_2^2\lambda_2 + \cdots + a_n^2\lambda_n}
\eeq
The relative change in squared distance will therefore be somewhere
between a maximum of $\omega\lambda_1$ and a minimum of
$\omega\lambda_n$, depending on the values of the $a_i$.  If the
$\lambda_i$ differ substantially, one would expect that chains started
with random $x$ and $x'$ would approach rapidly initially, but more
slowly later, once the chains are close together in the $v_1$ direction.

Unfortunately, once the chains are close enough that the second term
of (\ref{eq-change}) is substantial, the tendency for them to approach
closer will be largely lost.  Exact coalescence will then occur only
when the chains come together by chance.  

We can try to find roughly when the chains cease to systematically
approach by equating the expected decrease in the squared distance due
to the first term in (\ref{eq-change}), which is given by
(\ref{eq-edec}), with the expected increase due to the second term in
(\ref{eq-change}), which is
\beq
  E\Big[\, \delta^{T}\Sigma^{-1}\delta\ |(x-x')^{T}\Sigma^{-1}\delta|/2\,\Big]
\label{eq-einc}
\eeq
Finding the value of this exactly seems hard, but we can use it to
find how the distance of $x$ from $x'$ at the time when the chains
stop approaching scales with $\omega$ (which as defined above is
proportional to $w^2$).  Let $d^2$ be the squared magnitude of $x-x'$,
and assume that the direction of $x-x'$ at the relevant time is
independent of $\omega$, as should be true for sufficiently small
values of $\omega$.  We can then see that the expected increase in
distance from (\ref{eq-einc}) is proportional to $\omega^{3/2}d$,
while the expected decrease in distance from (\ref{eq-edec}) is
proportional to $\omega d^2$.  From this, one can see that the squared
distance at which the chains cease to systematically approach each
other should be proportional to $\omega$; hence the distance at which
the chains cease to approach each other is proportional to $w$.

\subsection{Demonstration of random-grid Metropolis for a multivariate 
normal}\label{sec-rgmvnex}\vspace*{-6pt}

The behaviour analysed in the previous section will be demonstrated
here using a nine-dimensional normal distribution for
$x_1,\ldots,x_9$.  The means of these variables are all zero.
Variables $x_7,\ldots,x_9$ have standard deviation 0.1, and are
independent of each other and of the other variables.  Variables
$x_1,\ldots,x_6$ have standard deviation one, and have correlation
$-0.199$ with each other.  This correlation is close to the value of
$-1/5$ at which the covariance matrix would become singular.  The
eigenvalues of $\Sigma^{-1}$ for this distribution are
$\lambda_1=200$, $\lambda_i=100$ for $i=2,\ldots,4$, and
$\lambda_i=0.834$ for $i=5,\ldots,9$.

\begin{figure}[p]
\centerline{\psfig{file=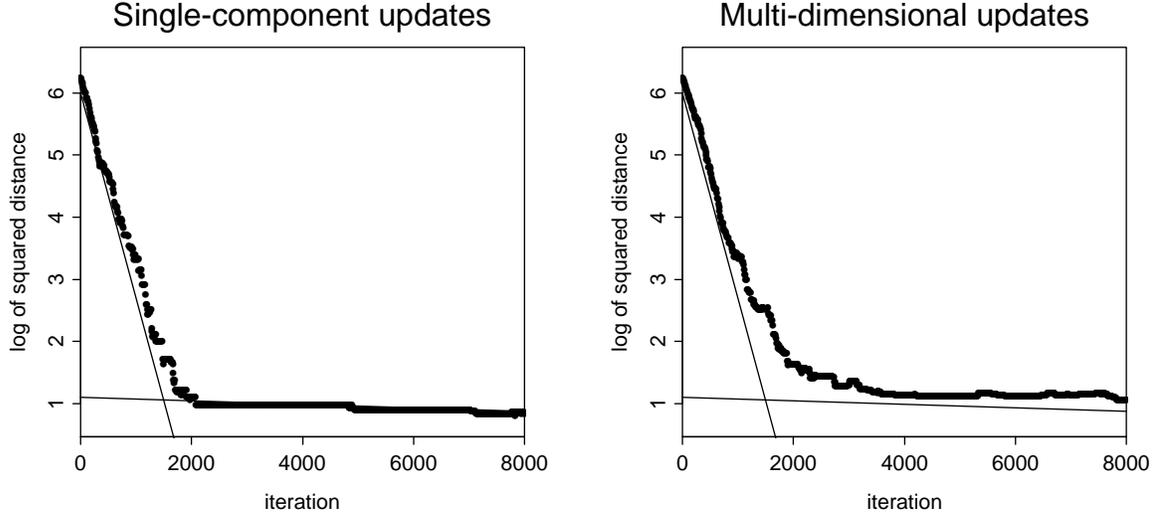}}

\vspace*{-15pt}

\caption[]{Initial approach of coupled chains using small 
stepsizes.  The plot on the left shows how the squared distance (with
metric given by $\Sigma^{-1}$) declines when
using single-component random-grid Metropolis updates; that on the right
shows the decline when using multi-dimensional updates.  The lines drawn 
on the plots have
slopes of $-\omega\lambda_2$ and $-\omega\lambda_9$, which are the 
rates of decline predicted from the analysis of
Section~\ref{sec-rgmvna} after the initial drop controlled by $\lambda_1$
and before the random influences become dominant.  The intercepts of the 
lines were chosen to fit the data.}\label{fig-rg1}
\end{figure}

\begin{figure}[p]
\centerline{\psfig{file=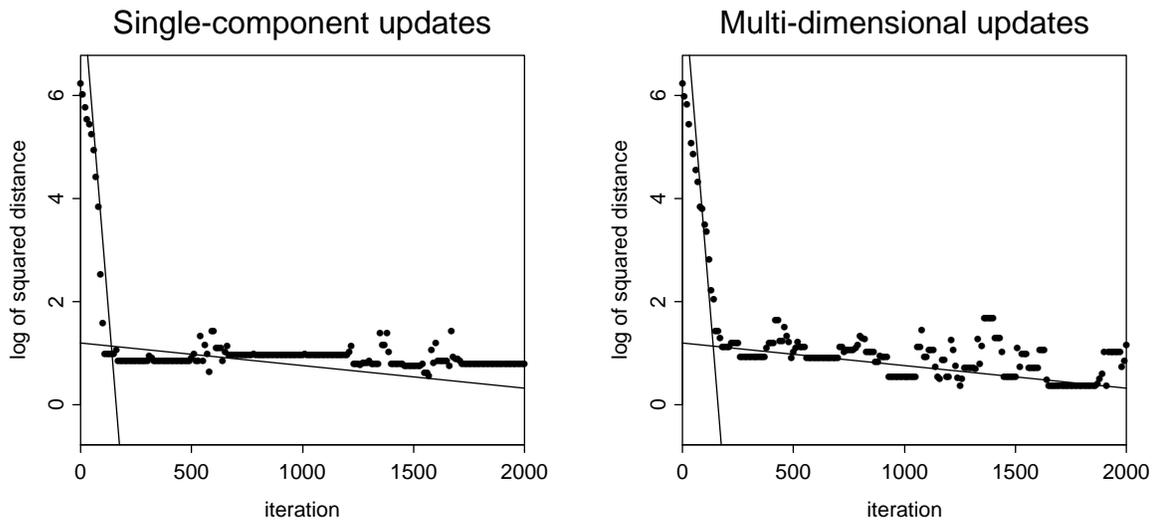}}

\vspace*{-15pt}

\caption[]{Initial approach of coupled chains using 
larger stepsizes.  The plots are analogous to those in 
Figure~\ref{fig-rg1}, except that the intercepts of the lines
were changed to fit this data.}\label{fig-rg2}
\end{figure}

Multi-dimensional random-grid Metropolis updates and single-component
random-grid Metropolis updates (of a component selected at random)
were both tried.  To begin, we can look at the results using quite
small stepsizes of $w=0.01$ for the multi-dimensional updates and
$w=0.03$ for the single-component updates.  Both stepsizes result in
$\omega\approx3.3\times10^{-5}$; the rejection rate is about 4.5\%.
With such small stepsizes, the assumptions of the analysis in the
previous section should be satisfied.

Figure~\ref{fig-rg1} shows simulations of two coupled chains with
these stepsizes, starting from initial states of
$x=(1.1,0.5,0,0,0,0,0.5,0.4,0.3)$ and
$x'=(-0.9,-0.5,0,0,0,0,-0.6,-0.4,-0.2)$.  The plots show the values of
the log of the squared distance of the two chains, given by
$(x-x')^{T}\Sigma^{-1}(x-x')$, at every tenth iteration, up to
iteration 8000.  The way the distance declines matches the predictions
of the analysis in the previous section.  Initially, the reduction in
the log of the squared distance averages about $\omega\lambda_2$ each
iteration, as expected if the reduction is dominated by $\lambda_2$ to
$\lambda_4$, which all have the value 100, only a bit less than the
value of $200$ for $\lambda_1$.  After about 2000 iterations, the rate
of decline changes, as expected, to about $\omega\lambda_9$.

With larger stepsizes of $w=0.04$ for multi-dimensional updates and
$w=0.12$ for single-component updates, the chains approach each other
faster, as shown in Figure~\ref{fig-rg2}.  For these stepsizes,
$\omega\approx5.3 \times 10^{-4}$, and the rejection rate is about
18\%.  The rates of decline in squared distance are mostly consistent
with the analysis of Section~\ref{sec-rgmvna}, but with these larger
stepsizes, the initial rate of decline is lowered, because the
probability of one chain accepting when the other rejects saturates at
one.  Also, random increases in distance are apparent fairly early
on.

\begin{figure}[t]
\centerline{\psfig{file=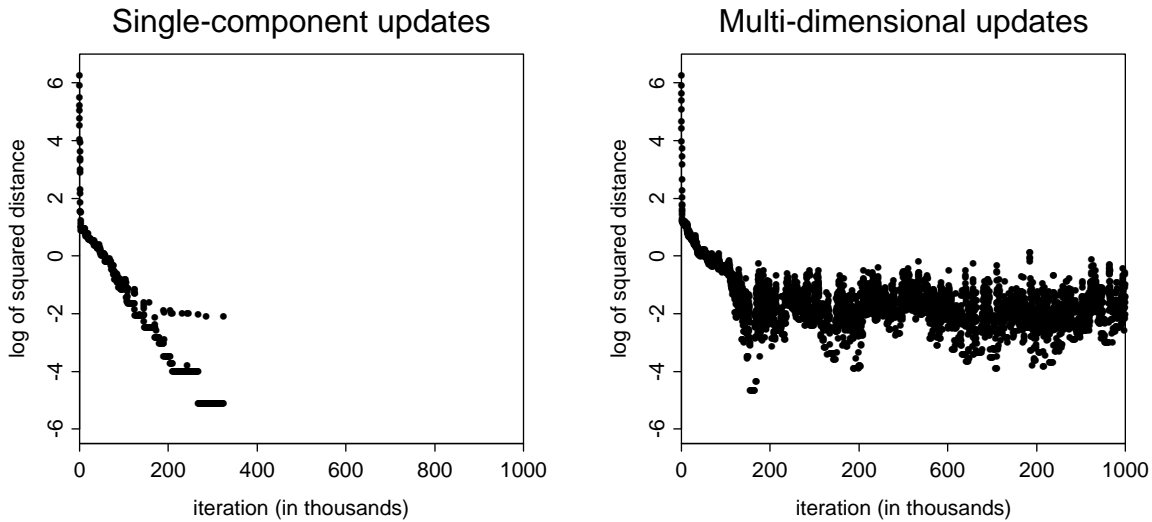}}

\vspace*{-15pt}

\caption[]{Coalescence of coupled chains using small stepsizes.  The 
plots extend those in Figure~\ref{fig-rg1}, but with only every hundredth 
iteration shown.  On the left, coalescence occurred
after 324,700 iterations; on the right,
it did not occur within one million iterations.}\label{fig-rg3}
\end{figure}

\begin{figure}[p]

\centerline{\psfig{file=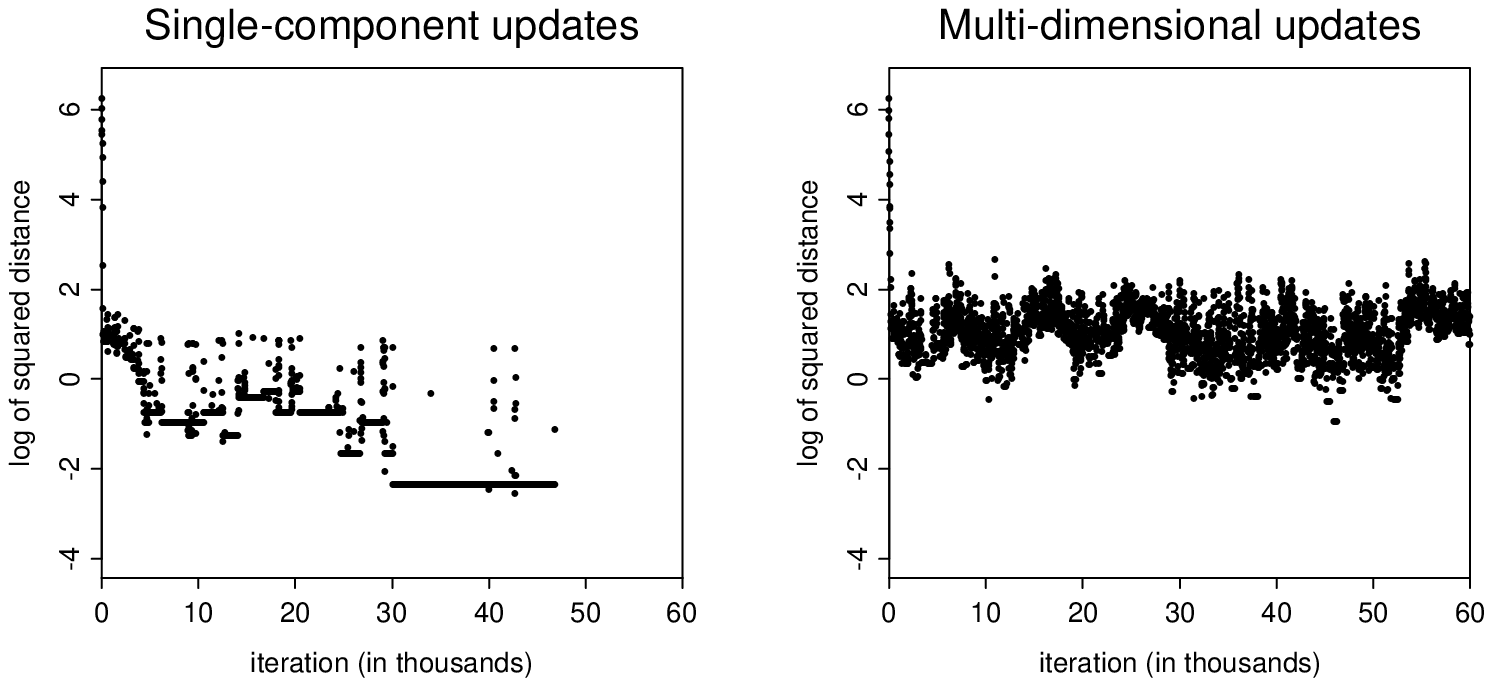}}

\vspace*{-8pt}

\psfig{file=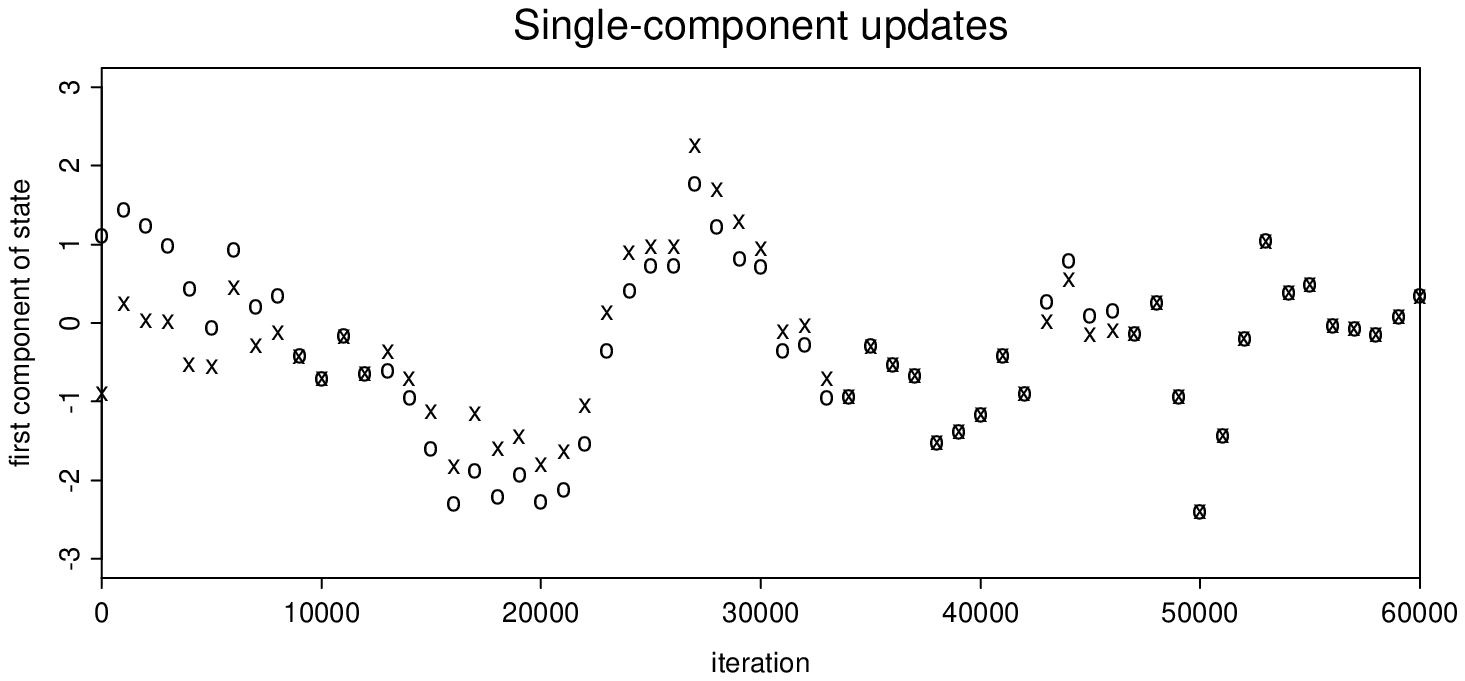}

\vspace*{-15pt}

\caption[]{Coalescence of coupled chains using larger stepsizes.  The 
top plots extend Figure~\ref{fig-rg2}.  The chains using single-component
random-grid updates coalesced after 46810 iterations; those using 
multi-dimensional updates did not coalesce within 60000 iterations. 
The bottom plot shows the first component of state for the pair of
chains that used single-component updates.  The states
of the two chains are plotted with ``x'' and ``o'' every 1000 
iterations.  Note that the two chains sometimes have exactly the
same value for this component prior to when they finally coalesce; at these
times, other components of the states still differ.}\label{fig-rg4}
\end{figure}

Longer runs show that after the initial approach, the distance between
the two chains reaches a quasi-stable random distribution, until such
time as the chains happen to coalesce by chance.  This is seen for the
smaller stepsizes in Figure~\ref{fig-rg3}, and for the larger
stepsizes in Figure~\ref{fig-rg4}.  The analysis of
Section~\ref{sec-rgmvna} suggests that the smaller stepsizes should
produce smaller values of the squared distance during this random
phase, in proportion to the values of $\omega$.  Here, the log of the
ratio of $\omega$ values is
$\log(5.3\times10^{-4}/3.3\times10^{-5})\approx2.8$, which does seem
to be roughly the difference in the upper end of the values seen in
Figures~\ref{fig-rg3} and~\ref{fig-rg4}.

Figure~\ref{fig-rg4} also includes a plot of the first component of
state for the chains using single-component updates, which indicates
that these chains move around the distribution in less time than it
takes for them to coalesce.  One would hope to find coupling schemes
that do better than this.  Coalescence using multi-dimensional updates
is even slower (with the same value for $\omega$), a phenomenon that
was confirmed with further runs (though there is considerable
variability in coalescence times for both methods).  The faster
coalescence when using single-component updates appears to be due to
the independence of $x_7$, $x_8$, and $x_9$ from the each other and
from the other components.  Because of this independence, once one of
these components becomes the same for the two chains, it will remain
the same thereafter, if all updates are made to single components,
since differing values for the other components will not affect its
conditional distribution.  In the run shown on the upper left of
Figure~\ref{fig-rg4}, $x_7$, $x_8$, and $x_9$ all coalesce by
iteration 110, leaving a six-dimensional rather than a
nine-dimensional problem.  This also explains why the upper end of
the quasi-stable distribution for distance is lower for single-component
updates than for multi-dimensional updates.

However, for much larger values of the stepsize, $w$,
multi-dimensional updates work better.  For example, with $w=0.64$,
performance is comparable to that seen in Figure~\ref{fig-rg4} for
single-component updates.  With this value for $w$, the acceptance
rate is very low (about 0.7\%), but the increased chances of
coalescence when proposals are accepted compensates.  In problems with
many more than the nine dimensions of this example, however, a
correspondingly large value for $w$ would likely lead to an extremely
small acceptance rate, so such a large-stepsize strategy cannot be
expected to work in general.

\section{Schemes that combine random-grid and other 
updates}\label{sec-schemes}\vspace*{-10pt}

Although coupled chains that use random-grid Metropolis updates are
capable of coalescencing exactly, we can see from
Figures~\ref{fig-rg3} and~\ref{fig-rg4} that the random-grid method is
not very efficient at producing exact coalescence of chains that have
approached closely.  This reflects a deficiency in the coupling
scheme, since these chains move to a nearly independent point in less
time than that required for them to coalesce exactly.  Furthermore,
even if they could be coupled more effectively, random-grid Metropolis
updates will not always be the best way of sampling from the
distribution of interest.

In this section, I explore a general strategy for combining
random-grid Metropolis updates with other types of updates to produce
Markov chains that sample efficiently and that are coupled so as to
produce exact coalescence in a time similar to that needed to move to a
nearly independent point.  I then discuss how well this strategy
works with standard Metropolis, Langevin, and Gibbs sampling updates.

\subsection{The general strategy}\label{sec-gstrategy}

The strategy that will be used for practical applications of circular
coupling consists of alternately performing two sorts of updates:
\begin{enumerate}
\item An update, or series of updates, that is designed to efficiently
      sample from the desired distribution, and that is coupled so as
      to cause chains to approach closer and closer.
\item A random-grid Metropolis update, which can lead to exact coalesce
      of chains that are already close together.
\end{enumerate}
The coupling scheme used for the updates in step~(1) may not be capable of 
achieving \textit{exact} coalescence, but if these updates have brought the 
chains close together, the random-grid update in step~(2) will have a good 
chance of bringing them together exactly.

A wide variety of Markov chain updates could be used in step~(1).
Mostly crucially, all such updates must be implemented in a way that
ensures that the same number of pseudo-random variates are generated
for an update regardless of the state of the chain --- if this is not
so, all subsequent updates will use different pseudo-random numbers in
different chains, destroying any possibility of coalescence.  Beyond
this, it is desirable for the updates in step~(1) to bring the coupled
chains closer together, though it is not essential that this occur for
every pair of states that the chains might be in, as long as it occurs
sufficiently often that there is a good chance that the chains will be
brought close enough together that the random-grid update of step~(2)
has a good chance of producing exact coalescence.  

In step (2), it is probably best to perform only a single random-grid
update for each component.  If the proposal in a random-grid update
is accepted by both chains, but fails to bring them together, a second
random-grid update using the same value of $w$ will certainly not do
so, since the states will differ by $2w$ or more in at least one
component, too far for them to coalesce in a single additional update
that uses the same value for $w$.  If an initial random-grid update is
rejected by one or both chains, a second update might produce
coalescence, but this possibility must be balanced against the
tendency of random-grid updates to move chains that have been brought
close together further apart.

There are two ways of performing a random-grid update of each
component --- by doing a single multi-dimensional random-grid update,
or by doing single-component random-grid updates for each component
in turn.  The following rough analysis indicates that for
high-dimensional problems a single multi-dimensional update is the
better choice.

Suppose that the components of the state in the two chains differ by
$d_1,\ldots,d_n$, where $n$ is the dimensionality.  Then the
probability of a randomly-placed grid with spacing $2w$ leading
to proposals in the two chains with the same value for some particular 
component is
\beq
   C_i(w) & = & \left\{\begin{array}{ll} 
                                      0        & \mbox{if $|d_i|\ge2w$}\\[3pt]
                                      1\!-\!|d_i|/2w & \mbox{if $|d_i|<2w$} 
                \end{array}\right.
\eeq
The probability that the proposals have the same values for all components,
and hence would produce exact coalescence if accepted, is therefore
\beq
  C(w) & = & \prod_{i=1}^n C_i(w) 
  \ \ =\ \ \left\{\begin{array}{ll} 
                                  0        & \mbox{if $\max |d_i|\ge2w$}\\[3pt]
          \prod\limits_i\, 1\!-\!|d_i|/2w  & \mbox{if $\max |d_i|<2w$} 
           \end{array}\right.
\label{eq-Cw}
\eeq
If the $d_i$ are much smaller than $w$, this can be approximated by
\beq
  \log C(w) & \approx & - {1 \over 2w} \sum_{i=1}^n |d_i|
               \ \ =\ \ - n \bar d / 2w
\label{eq-logCw}
\eeq

To obtain the probability of exact coalescence, we must multiply $C(w)$
by the probability that both chains accept a single multi-dimensional
random-grid update, or by the product of the acceptance probabilities
for all the single-component random-grid updates.  (I'll assume, as
seems reasonable, that acceptance is nearly independent of whether the
proposals would lead to coalescence if accepted.)  The value of $w$
affects the acceptance probability and proposal coalescence probability
oppositely, necessitating a trade-off.

For acceptance
probabilities not far from one, the log of the acceptance probability
for a single-component update will typically be roughly $-Kw$, for some
constant $K$, and the log of the probability that the updates to all
$n$ components will be accepted will be $-n \bar Kw$.  The optimal choice
for $w$ with single-component updates maximizes the log of the probability
of actual coalescence, which is
\beq
  A_s(w) & = & -n \bar K w - n \bar d / 2w
\eeq
The maximum occurs at $w_s=\sqrt{\bar d/2\bar K}$, where
$A_s(w_s)=-n\sqrt{2 \bar d \bar K\rule[-1pt]{0pt}{10pt}}$.  We would like 
$A_s(w_s)$ to have magnitude of order one, in which case we must have 
$\bar d \sim n^{-2}$, and $w_s \sim n^{-1}$.  Note that with this scaling 
for $w$, the probability that the proposals for all components will be 
accepted will remaining of order one as $n$ increases.

In contrast, for a multi-dimensional update, the log of the acceptance
probability will be roughly $-\sqrt{n}\bar Kw$ (Roberts, \textit{et al.}
1997).  The optimal $w$ will maximize
\beq
  A_m(w) & = & -\sqrt{n}\bar Kw - n \bar d /2w
\eeq
The maximum occurs at $w_m=n^{1/4}\sqrt{\bar d/2\bar K}$, where 
$A_m(w_m)=-n^{3/4}\sqrt{2\bar d \bar K\rule[-1pt]{0pt}{10pt}}$.  
For the magnitude 
of $A_m(w_m)$ to be of order one, we must have $\bar d \sim n^{-3/2}$, and 
$w_m \sim n^{-1/2}$.  Note that with this scaling for $w$, the acceptance 
probability will remain of order one as $n$ increases.

We can conclude that when $n$ is large, multi-dimensional random-grid
updates will be preferred to single-component updates in step~(2),
since they will have a good chance of producing exact coalescence with
a larger value for $\bar d$, which will be easier to achieve in
step~(1).  Exceptions to this may occur when some components of state
are independent of others, in which case exclusive use of
single-component updates in both steps (1) and (2) can result in these
components coalescing independently, invalidating the assumption above
that coalescence requires that a single proposal or sequence of
proposals lead to coalescence of all components simultaneously,
starting from a state in which none of the components had coalesced.

It would be possible to use random-grid Metropolis updates in both
step~(1) and step~(2), likely with different values for $w$.  There
would then be at least a small chance of exact coalescence occurring in
step~(1).  I will not examine this possibility here, however, since
using more general updates for step~(1) provides greater flexibility
when chosing an efficient sampling scheme for the distribution of
interest.  

In the sections below, I consider the use in step~(1) of standard
random-walk Metropolis updates, Langevin updates, and Gibbs sampling
updates.  I first investigate in some detail how well standard
Metropolis updates work, because this method is well-known and
widely-used.  It will turn out, however, that although standard
Metropolis updates can be used in this coupling scheme, they are
rather inefficient, and difficult to tune.  Some readers may therefore
wish to skip immediately to Section~\ref{sec-clang}, where Langevin
updates are shown to work much better.

\subsection{Coupled Metropolis updates}\label{sec-cmet}\vspace*{-6pt}

Standard random-walk Metropolis updates use a symmetrical proposal
distribution centred on the current state.  Such a proposal
distribution can be viewed as adding a random offset to the state, and
updates based on these proposals can be coupled by simply using the
same random offset in different chains, as well as using the same
random number to make the acceptance decision.  The uniform proposal
of equation~(\ref{eq-prop}), coupled as in
equation~(\ref{eq-couple0}), is one example, which can be generalized
to multi-dimensional proposals.  It is also common for the offset of
the proposed state to come from a zero-mean normal distribution.  For
multi-dimensional updates, the covariance matrix for such proposals is
often diagonal, since obtaining detailed information about the
distribution that would allow a better choice may be difficult, or the
dimensionality may so large as to make general matrix operations
costly.

The analysis of standard random-walk Metropolis updates coupled by
using the same offsets in different chains is similar to that for
random-grid Metropolis updates, presented in Section~\ref{sec-rgmvna},
for sampling from multivariate normal distributions.  The analysis of
standard Metropolis updates is simpler, however, since the
complication of the four types of random-grid updates is absent.
Instead, the offset of the proposed state from the current state will
always be the same in all chains, and we can analyse all updates in
the same way as type~(3) random-grid updates were analysed.

If variables are updated one at a time, at random, using a normal
proposal distribution with standard deviation $\sigma$, the expected
squared change to a coordinate will be $\omega = \sigma^2/n$, where
$n$ is the number of dimensions.  Multi-dimensional updates using a
proposal distribution with covariance matrix $\sigma^2I$ will of
course give $\omega=\sigma^2$.

\begin{figure}[p]
\centerline{\psfig{file=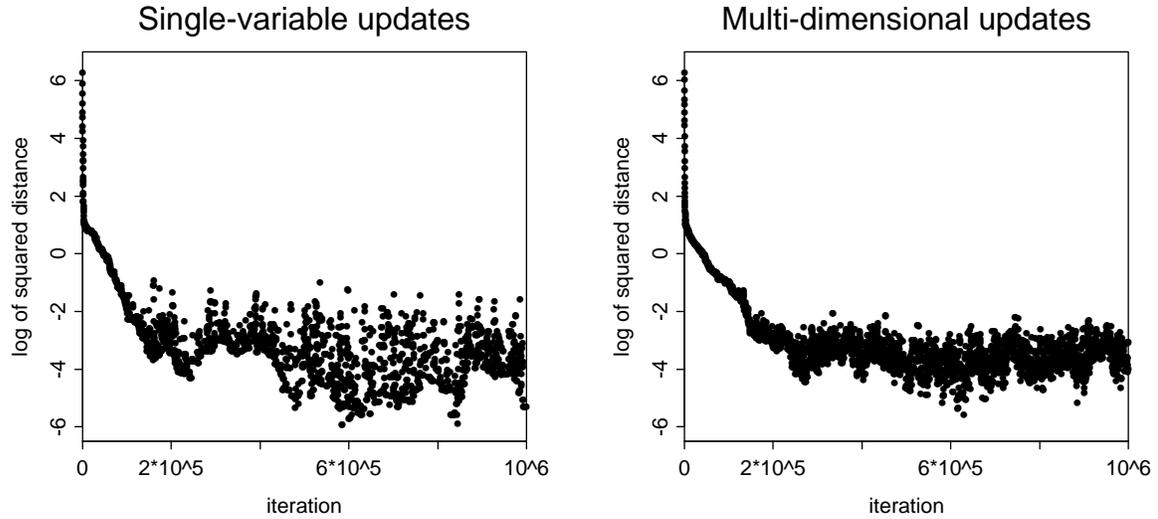}}

\vspace*{-15pt}

\caption[]{Approach of chains using only standard Metropolis updates
based on a normal proposal distribution with small stepsizes, as
measured by the log of the squared distance under the metric defined by
$\Sigma^{-1}$.  Compare with the behaviour using random-grid updates shown in
Figure~\ref{fig-rg3}.}\label{fig-sm1} 
\end{figure}

\begin{figure}[p]
\centerline{\psfig{file=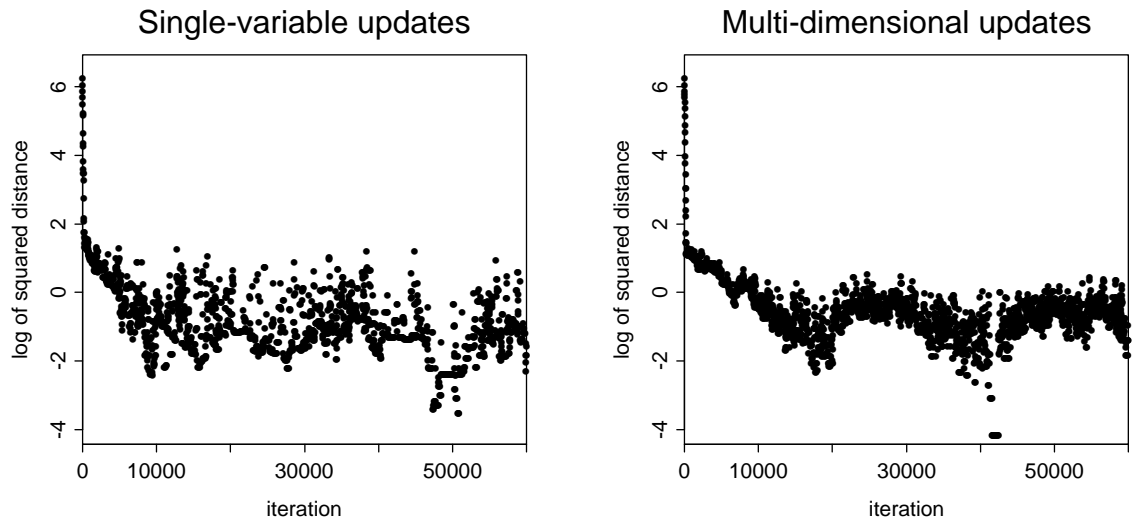}}

\vspace*{-15pt}

\caption[]{Approach of chains using only standard Metropolis updates based
on a normal proposal distribution with 
larger stepsizes.  Compare with the behaviour using random-grid updates shown
in Figure~\ref{fig-rg4}.}\label{fig-sm2}
\end{figure}

Figures~\ref{fig-sm1} and~\ref{fig-sm2} illustrate that chains using
standard Metropolis updates approach each other in the same way as
chains using random-grid Metropolis updates, for the nine-dimensional
normal distribution described in Section~\ref{sec-rgmvnex}.  The right
plot in Figure~\ref{fig-sm1} shows the results that are obtained
when updating single components chosen at random with a normal
proposal distribution for which $\sigma=0.017$; the plot on the left
shows the results obtained using multi-dimensional updates with a
spherical normal proposal distribution for which $\sigma=0.0058$.  In
both cases, $\omega\approx3.3\times10^{-5}$, the same as in
Figure~\ref{fig-rg3}.  Larger stepsizes of $\sigma=0.069$ and
$\sigma=0.023$, giving $\omega\approx5.3\times10^{-4}$, were used in
Figure~\ref{fig-sm2}, corresponding to the random-grid results in
Figure~\ref{fig-rg4}.  The rejection rates of the standard Metropolis
methods were similar to those of the corresponding random-grid
methods.

The behaviour seen with these standard Metropolis updates is both
qualitatively and quantitatively similar to that seen with random-grid
Metropolis updates.  An exception visible in the figures is that when
using standard multi-dimensional Metropolis updates, the distance at
which random increases in distance between chains begin to occur is
less than for the corresponding runs with random-grid Metropolis
updates.

A fundamental difference is that chains coupled using random-grid
updates eventually coalesce exactly.  In contrast, it is easy to see
that when two chains using standard Metropolis updates with a
continuous proposal distribution are coupled by using the same
offsets, the probability of exact coalescence after any finite number
of iterations is zero, since it can occur only if the random proposal
offset exactly matches the difference between the states of the two
chains (and one then accepts while the other rejects).

\begin{figure}[t]

\psfig{file=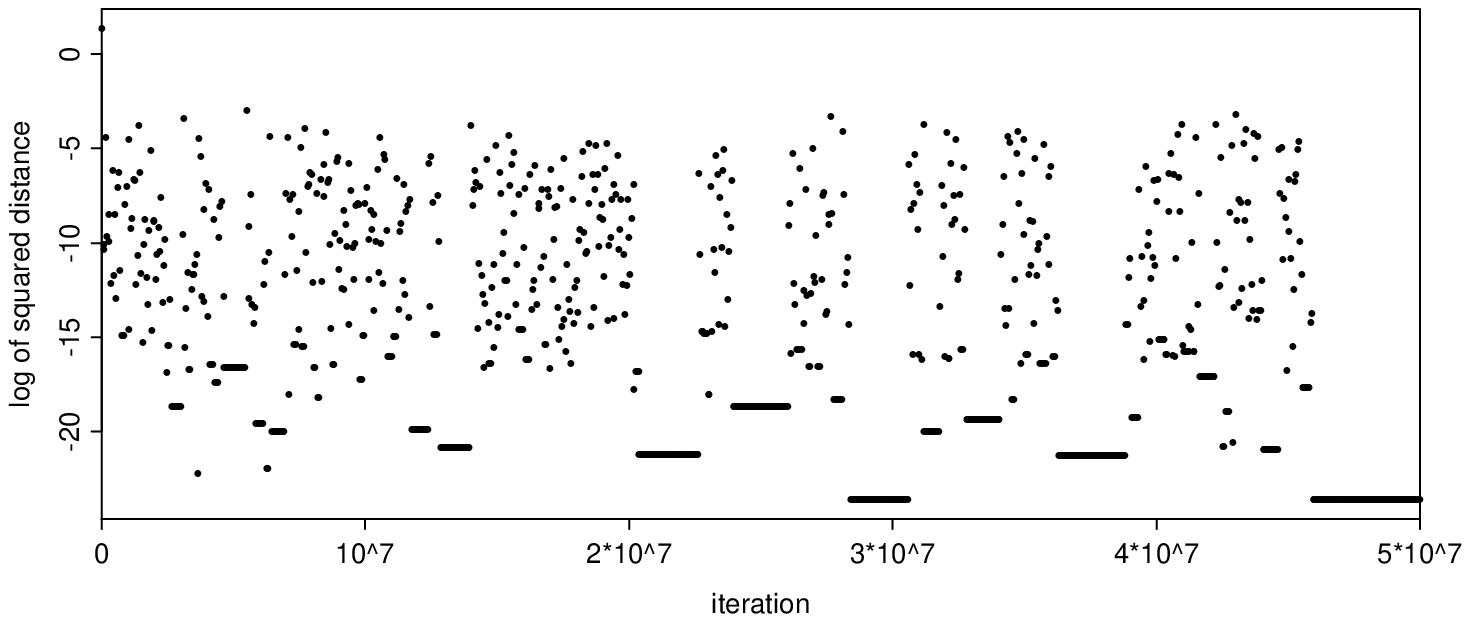}

\vspace*{-15pt}

\caption[]{Coupling of standard Metropolis updates sampling from a 
univariate standard normal distribution.  The proposal distribution was normal
with standard deviation 0.1.  The plot shows the log of the squared distance
between two chains started at $+1.1$ and $-0.9$ at every 50000 iterations
of a run of length $5\times10^7$.}\label{fig-sm3}
\end{figure}

For this nine-dimensional example, the distance between coupled chains
using standard Metropolis updates appears to converge to a
non-degenerate stationary distribution.  In one dimension, however,
coupled chains spend ever-increasing amounts of time at ever-smaller
distances.  This is illustrated in Figure~\ref{fig-sm3}, which shows
two chains sampling from a standard univariate normal distribution.
In $n$ dimensions, the probability that a standard Metropolis update
will result in two chains that were moderately close approaching to a
small distance between $r$ and $r+dr$ will be proportional to
$r^{n-1}dr$, based on the volume of a shell of this radius.  Once
there, these two chains will remain at that distance until one accepts
at the same time as the other rejects.  The probability of this
happening when sampling from a normal distribution is proportional to
$r$, so the expected time spent at this distance before jumping back
to a moderate distance will be proportional to $1/r$.  The fraction of
time spent at a distance between $r$ and $r+dr$ will therefore be
proportional to $r^{n-2}dr$, which is $dr/r$ when $n=1$.  Since the
integral of $dr/r$ diverges in the vicinity of zero, an
ever-increasing amount of time will be spent with the chains in
ever-closer states.  This will not happen for $n>1$, however, unless
the problem is essentially one-dimensional, due to the components
being independent, with single-component updates being used.  The
phenomenon is probably of no practical help in achieving exact
coalescence even in one dimension.

Instead, we can follow the two-step strategy outlined in
Section~\ref{sec-gstrategy}, combining standard Metropolis updates
designed to bring chains close together (step (1)) with a random-grid
Metropolis update designed to produce exact coalescence once the
chains are close (step (2)).  As discussed in that section, exact
coalescence can occur in step~(2) only if the separation of the two
chains for all components is less than $2w$, where $w$ is the
parameter of the random-grid update.  Exact coalescence will occur
with reasonably high probability only if in addition the average
separation, $\bar d$, is small enough that a coalescent proposal is
likely, and if the probability of accepting this proposal is both
chains is fairly high.

\begin{figure}[p]

\centerline{\psfig{file=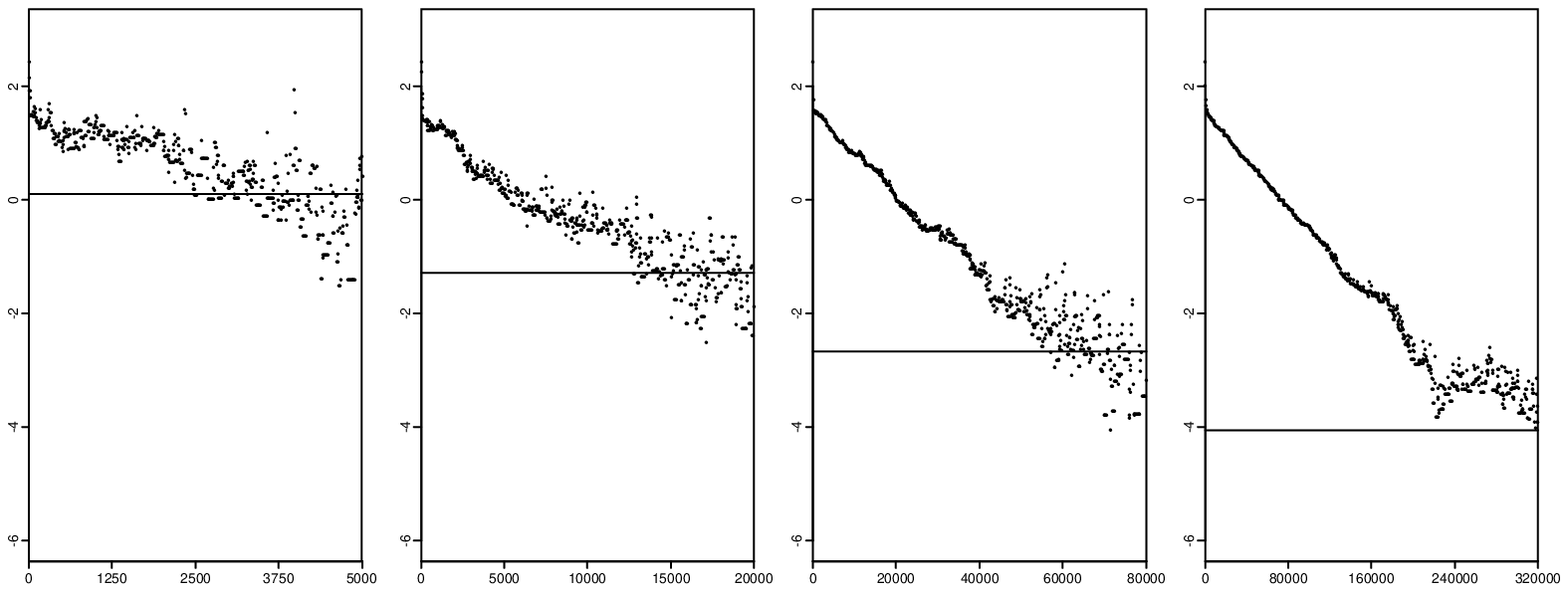}}

\vspace*{-10pt}

\hspace*{-30pt}\hfil
\makebox[1in][r]{$\sigma=0.04$}\hfil\hfil
\makebox[1in][r]{$\sigma=0.02$}\hfil\hfil
\makebox[1in][r]{$\sigma=0.01$}\hfil\hfil
\makebox[1in][r]{$\sigma=0.005$}\hfil

\psfig{file=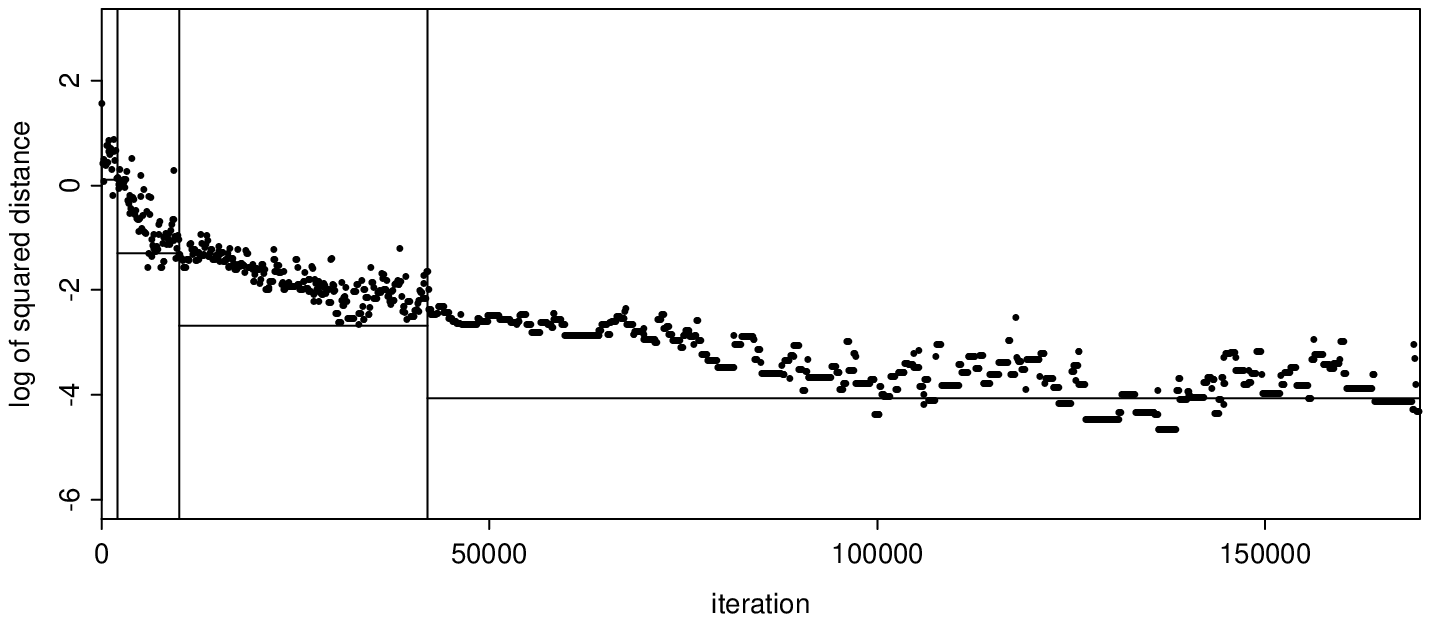}

\vspace*{-15pt} \caption[]{Coupling when sampling from the
nine-dimensional normal example using standard multi-dimensional
Metropolis updates, with varying stepsizes.  The top four plots show
the declines in the log of the squared distance (as defined by
$\Sigma^{-1}$) for four values of the proposal standard deviation,
$\sigma$.  Note the differing horizontal scales.  The horizontal lines
drawn on these plots indicate the approximate centres of the
stationary distributions for each $\sigma$.  They differ by the logs
of the ratios of the $\sigma^2$ values (ie, by $\log(4)$), as expected
from the theoretical analysis, with their overall vertical position
fit to the data by eye.  The lower plot shows how the log of the
squared distance declines during a run in which these four values for
$\sigma$ are used in succession, with the switching points marked by
vertical lines, and horizontal lines drawn as in the top
plots.}\label{fig-smi1}

\end{figure}

Chains can be brought close together more efficiently in step~(1) if
the stepsize ($\sigma$) in the standard Metropolis updates is varied,
starting with a large value that quickly brings the chains fairly
close together, then switching to a smaller value to bring them
somewhat closer, and so forth, until they are close enough for exact
coalescence to be likely in step~(2).  This is illustrated in
Figure~\ref{fig-smi1}, for the nine-dimensional normal distribution
described in Section~\ref{sec-rgmvnex}, using standard
multi-dimensional Metropolis updates.  The starting states were the
result of many multi-dimensional Metropolis updates using the fairly
large stepsize of $\sigma=0.1$.  By switching to successively smaller
values for $\sigma$, starting with $\sigma=0.04$, the squared distance
between the chains is reduced in 170000 iterations to a level that
would have taken over twice as many iterations if the smallest value
for $\sigma$ had been used from the beginning.  The pattern of decline
followed by random movement seen in the upper plots in
Figure~\ref{fig-smi1} illustrates that switching to smaller values of
$\sigma$ is a generally desirable strategy, though the optimal time to
switch is not obvious.

If, as in Figure~\ref{fig-smi1}, each stage uses a stepsize, $\sigma$,
that is a factor of two smaller than in the previous stage, we can
expect each stage to reduce the log of the squared distance by
$\log(4)$ before random increases in distance again become a large
factor.  Since the rate of decline in the log of the squared distance
is proportional to $\sigma^2$, accomplishing this will take four times
as many iterations as were needed in the previous stage.

We can determine the effort needed to reach a given distance by
methods similar to this as follows.  Suppose we go through stages
indexed by $s=1,2,\ldots$, using stepsizes in each stage of
$\sigma(s)=\sigma_0 e^{-as}$, starting with states a distance $D_0$
apart.  The minimum squared distance, $D(s)$, achievable in stage $s$
before random effects become large will be proportional to $\sigma^2$,
giving $D(s)=D_0 e^{-2as}$ or $\log D(s) = \log D_0 - 2as$.  The rate
of decline in $\log D(s)$ will be proportional to $\sigma^2$, giving a
rate in stage $s$ of $R(s)=R_0 e^{-2as}$.  The number of iterations
that must be spent in stage $s$ in order to reduce $\log D(s)$ by $2a$
will therefore be $\tau(s) = 2a/R(s) = (2a/R_0) e^{2as}$.  The total
number of iterations up to the completion of stage $s$ will be 
\beq
  T(s) & = & \sum_{i=1}^{s} \tau(i) 
  \ \ =\ \ {2a \over R_0} \sum_{i=1}^{s} e^{2ai} 
  \ \ =\ \ {2a \over R_0} {e^{2a(s+1)}-e^{2a} \over e^{2a} -1}
\eeq
To achieve some desired squared distance, $D_*$, will require $s_* = 
(\log D_0 - \log D_*)/(2a)$ stages, which will take a total number
of iterations of
\beq
  T(s_*) & = & {2a \over R_0} {e^{2a} \over e^{2a}-1} \left( {D_0 \over D_*} 
                - 1 \right)
\eeq
This becomes smaller as $a$ approaches zero, corresponding to changing 
$\sigma$ more and more frequently.  The limiting value as $a\rightarrow0$
is 
\beq
  T_* & = & {1 \over R_0} \left( {D_0 \over D_*} - 1 \right)
\label{eq-metTD}\eeq
Hence, to bring the chains close together will require time that is
inversely proportional to the square of the desired distance, if 
the optimal strategy of varying $\sigma$ is used.  If instead
we use the single stepsize of $\sigma(s_*)$, achieving
a squared distance of $D_*$ will require reducing $\log D$ by
$2as_*$.  The number of iterations required for this will be 
\beq
  {2as_* \over R_0}\ e^{2as_*} & = & 
  {\log (D_0/D_*) \over R_0} \ {D_0 \over D_*}
\eeq
Comparing with equation~(\ref{eq-metTD}), we see that varying $\sigma$
is advantageous, though the benefit is not huge.

\begin{figure}[p]

\centerline{\psfig{file=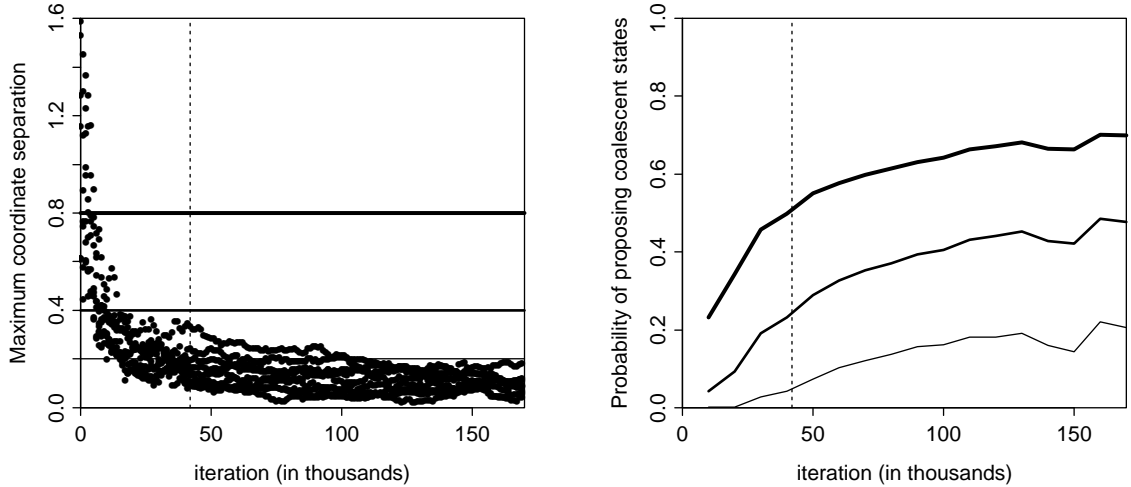}}

\vspace*{-20pt}

\caption[]{Maximum separation and probability of a coalescent proposal.
The left plot shows the maximum separation of components for the run shown 
in the bottom plot of Figure~\ref{fig-smi1} and eight replicatons with 
different random seeds.  Three values for $w$ in a random-grid update are 
considered: $w=0.1$, $w=0.2$, and $w=0.4$.  The corresponding values for $2w$, 
marking the point where coalescence is possible, are plotted as horizontal
lines on the left.  The nine runs are used to estimate the probabilities of 
a coalescent proposal with these values of $w$, plotted on the 
right, with thicker lines for larger values of $w$.}\label{fig-smi2}

\end{figure}

\begin{figure}[p]

\centerline{\psfig{file=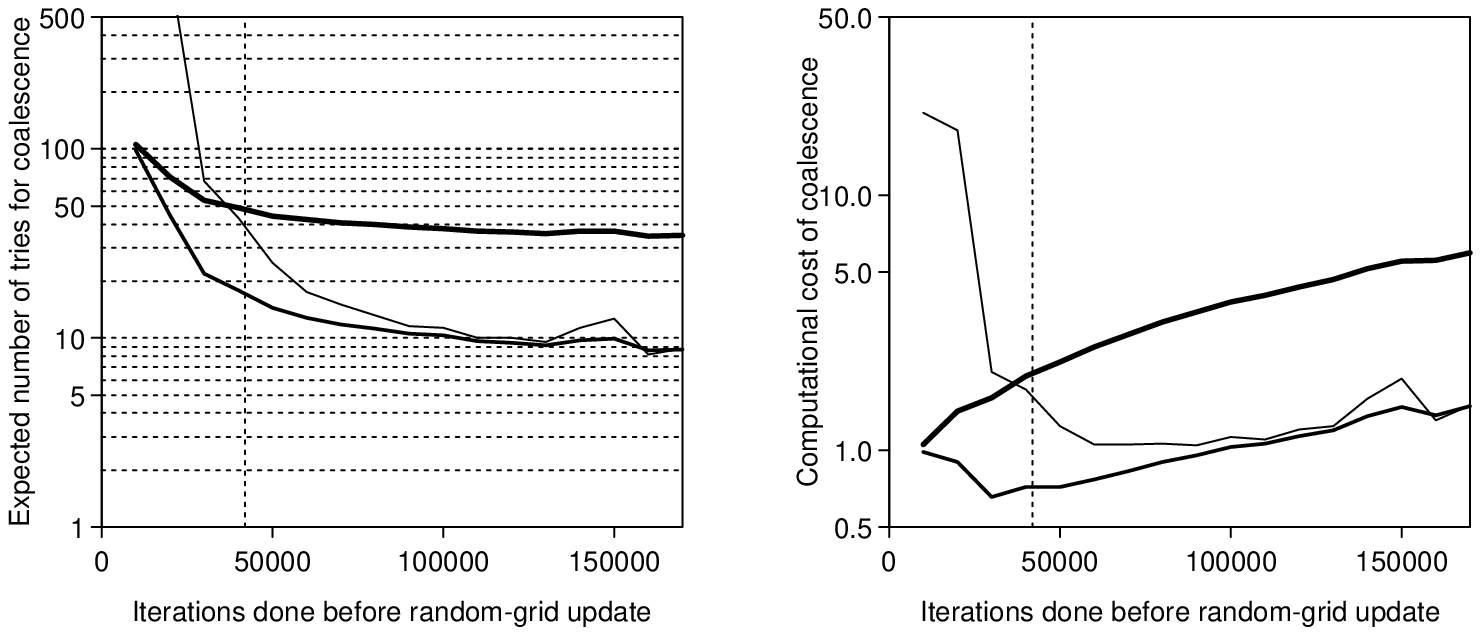}}

\vspace*{-15pt}

\begin{center}\small

Acceptance rates:\ \
 $w\!=\!0.1: 55\%,\ \ 
  w\!=\!0.2: 24\%,\ \ w\!=\!0.4: 4.1\%$

\end{center}

\caption[]{Effectiveness of coupling with different
numbers of iterations before a random-grid update and different values for $w$,
based on the estimated probabilities of coalescent proposals shown in
Figure~\ref{fig-smi2}.  The acceptance rates for the values of $w$ considered
are shown below the plots.  The
plot on the left shows the expected number of random-grid updates required
to obtain coalescence.  The plot on the right shows the corresponding 
computational cost, in millions of standard Metropolis iterations
before coalescence.
}\label{fig-smi3}

\end{figure}

More directly relevant than the squared distance is the maximum
separation of states in any coordinate direction, since coalescence
from a random-grid update is not possible if this is greater than
$2w$.  This maximum separation was investigated using nine runs like
the one shown at the bottom of Figure~\ref{fig-smi1}.  Once the
distance between the chains becomes small, the maximum separation was
found to be roughly proportional to the distance.  The maximum
separation of coordinates during the nine runs is plotted on the left
in Figure~\ref{fig-smi2}.  On the right in Figure~\ref{fig-smi2} is
the probability that a multi-dimensional random-grid proposal, or
sequence of single-component proposals, would result in the proposed
new states in both chains being the same, for three values of $w$,
found using equation~(\ref{eq-Cw}), and the distribution of component
separations seen in the nine runs.

To find the optimal value for $w$, we must also account for the
probability that a random-grid update that would lead to coalescence
if accepted by both chains actually is accepted by both.  Once the
chains are close, acceptance in one chain should usually coincide with
acceptance in the other, and acceptance should be nearly independent
of whether the proposals in the two chains would lead to coalescence.
With these assumptions, the probability of the chains coalescing is
the product of the probability of a coalescent proposal and the
acceptance rate.  The reciprocal of this is the expected number of
tries before coalescence occurs.  This is plotted for three values of
$w$ on the left in Figure~\ref{fig-smi3}, using the proposal
coalescence probabilities from the plot on the right of
Figure~\ref{fig-smi2}, and acceptance rates obtained empirically, and
shown in the figure.  Multiplying the expected number of tries by the
number of standard Metropolis iterations required in step~(1) for each
try gives a measure of the computational cost for achieving coalescence,
plotted on the right of Figure~\ref{fig-smi3}.

I tested these predictions by running the basic circularly-coupled
simulation procedure of Section~\ref{sec-proc}, with $N=100$ Markov
chain transitions, each consisting of steps~(1) and~(2) of the
procedure of Section~\ref{sec-gstrategy}.  Step~(1) consisted of
$170000$ standard multi-dimensional Metropolis iterations with varying
$\sigma$, as in the run shown in the bottom of Figure~\ref{fig-smi1}.
Step~(2) consisted of one multi-dimensional random-grid update.  For
each of the values of $w$ considered in Figures~\ref{fig-smi2}
and~\ref{fig-smi3}, nine runs were done, all starting from the state
$x=0$, and the number of iterations before the wrapped-around chain
coalesced with the original chain was recorded.
Runs were also done in which step~(1) consisted of only $42000$
standard Metropolis updates, corresponding to stopping after the
updates with $\sigma=0.01$.  Finally, runs were done with $170000$
standard multi-dimensional Metropolis iterations followed by one
sequence of single-component random-grid updates.  

\begin{table}

\begin{tabular}{|l|rrlr|rrlr|rrlr|} \hline
 & \multicolumn{4}{c|}{\small multi-dimen. random grid}
 & \multicolumn{4}{c|}{\small multi-dimen. random grid} 
 & \multicolumn{4}{c|}{\small single-comp. random grid} \\
 & \multicolumn{4}{c|}{\small 170000 iterations}      
 & \multicolumn{4}{c|}{\small 42000 iterations}      
 & \multicolumn{4}{c|}{\small 170000 iterations} \\ \hline
 $w=0.1$
 &  8$\!\!$ & (5, & $\!\!\!\!$16) &  9 
 & 34$\!\!$ & (21,& $\!\!\!\!$66) & 41 
 & 12$\!\!$ & (7, & $\!\!\!\!$23) & 19  \\
 $w=0.2$
 &  7$\!\!$ & (4, & $\!\!\!\!$13) &  9 
 & 15$\!\!$ & (9, & $\!\!\!\!$28) & 17 
 & 38$\!\!$ & (23,& $\!\!\!\!$77) [1] & 31  \\
 $w=0.4$
 & 18$\!\!$ & (11,& $\!\!\!\!$34) & 35 
 & 73$\!\!$ & (42,& $\!\!\!\!$168) [3] & 48 
 &841$\!\!$ & (281,& $\!\!\!\!$16396) [8]$\!\!\!\!$ & 572 \\ \hline
\end{tabular}

\caption[]{Comparison of observed and predicted coalescence
times.  For each type of run and each value of the random grid
stepsize, $w$, the posterior mean (based on an exponential model)
of the mean time to coalescence is given, followed by the 90\%
posterior interval (with symmetric 5\% tails), and finally 
the theoretical prediction.  The numbers in square brackets are
the numbers of censored observations (ie, runs that did not coalesce within
100 steps), out of nine total.}\label{tbl-mres}

\end{table}

The results are shown in Table~\ref{tbl-mres}.  The coalescence times
--- ie, the number of random-grid updates before coalescence, each
preceded by a series of standard Metropolis updates --- were modeled
as being exponentially distributed, censored at 100 (the length of a
run).  The mean coalescence time was given an improper prior with
density proportional to the reciprocal of the mean.  The table shows
the posterior mean (equal to the sample mean if all runs coalesced
before 100 iterations), along with a 90\% posterior interval.  This is
followed by the theoretical prediction for the mean coalescence time,
as shown in Figure~\ref{fig-smi3}, or computed in an analogous manner
for the single-component random grid updates.  (The probability of the
single-component random-grid proposals all being accepted was found
empirically to be 26\%, 6.7\%, and 0.25\% for $w=0.1$, $w=0.2$, and
$w=0.4$.)

There are no apparent conflicts between the observations and the theory,
showing that the various assumptions made hold at least approximately for 
this problem.

The overall conclusion I reach from this investigation is that
although standard Metropolis updates can be used to achieve
coalescence in a circularly-coupled simulation, doing so is rather
laborious, both in the amount of computer time required and in the
amount of human time needed to find an appropriate schedule of changes
to $\sigma$ and an appropriate choice of $w$.  One should note that
much of the time needed to achieve coalescence is spent doing updates
with stepsizes that are sub-optimally small from the point of view of
moving efficiently around the distribution.  Furthermore, as discussed
in Section~\ref{sec-gstrategy}, in high-dimensional problems, the
distance between chains will need to be made quite small in step~(1)
if step~(2) is to have a good chance of producing exact coalescence.
Achieving this with standard Metropolis updates is possible, using the
strategy of varying $\sigma$, but the scaling of the time needed for
the chains to approach within a given distance, given by
equation~(\ref{eq-metTD}), is rather poor.  Fortunately, there are
other Markov chain sampling methods that can be coupled much more
effectively.

\subsection{Coupled Langevin updates}\label{sec-clang}\vspace*{-6pt}

Langevin methods are applicable to distributions on continuous state
spaces for which the gradient of the probability density can be
computed.  Defining $E(x)=-\log \pi(x)$, an uncorrected Langevin update
changes the current state, $x_t$, to the next state
computed as follows:
\beq
  x_{t+1} & = & 
   x_t \ -\ (\epsilon^2/2) \nabla E(x)\ +\ \epsilon u_t \label{eq-ulang}
\eeq
where $u_t$ is a standard normal random variate, and $\epsilon$ is an
adjustable stepsize parameter.  The stationary distribution
of this Markov chain is not exactly $\pi$, but it approaches $\pi$
as $\epsilon$ approaches zero.  (Smaller values of $\epsilon$ will 
of course require longer runs.)

The corrected Langevin procedure, introduced by Rossky, Doll, and
Friedman (1978), produces the exactly correct stationary distribution
for any stepsize, $\epsilon$, by treating the right side of
equation~(\ref{eq-ulang}) as a Metropolis-Hastings proposal, which is
accepted or rejected based on ratios of probability densities and of
proposal densities (Hastings 1970).  The corrected
Langevin algorithm can also be seen a special case of more general
algorithms based on the ``leapfrog'' discretization of Hamiltonian
dynamics (Duane, \textit{et al} 1987; Neal 1996a, Section 3.1).  In this
formulation, we introduce a ``momentum'' vector, $p$, of the same
dimension as $x$, and define $H(x,p) = E(x) + |p|^2/2$.  We then
simulate a Markov chain that samples from the distribution for $(x,p)$
with density proportional to $\exp(-H(x,p))$, whose marginal
distribution with respect to $x$ is $\pi(x)$.

One iteration of this formulation of the Langevin method changes the 
state, $(x,p)$, as follows:\vspace*{-5pt}
\begin{enumerate}
\item Replace $p$ with an independent draw from the normal distribution
      with mean zero and covariance $I$.
\item Set $p' \ =\ p - (\epsilon/2) \nabla E(x)$
\item Set $x^* \ =\ x_t + \epsilon p'$
\item Set $p^* \ =\ p' - (\epsilon/2) \nabla E(x^*)$
\item Accept $(x^*,p^*)$ as the new state with probability 
      $\min\,[1,\,\exp(-H(x^*,p^*)+H(x,p))]$.  Otherwise, keep $x$ unchanged
      but negate $p$.\vspace*{-5pt}
\end{enumerate}
Step (1) and steps (2) to (5) can separately be shown to leave invariant
the distribution with density proportional to $\exp(-H(x,p))$.  The
combination therefore leaves this distribution invariant as well.  For
good performance, $\epsilon$ should be chosen to be small
enough that step (5) usually accepts, but not much smaller than is necessary
to achieve this.

\begin{figure}[t]

\centerline{\psfig{file=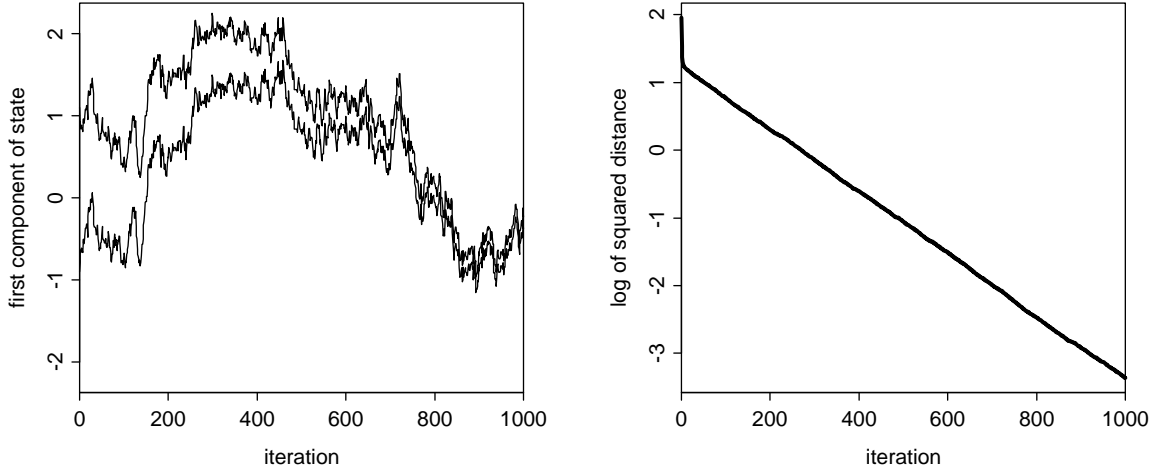}}

\vspace*{-20pt}

\caption[]{Sampling from a multivariate normal distribution using
coupled Langevin chains (with $\epsilon=0.08$) started from different
initial states.  The plot on the left shows the first component of
state for the two coupled chains.  The plot on the right show the log
of the squared Euclidean distance between the states of the two
chains.}\label{fig-lang1}

\end{figure}

To couple Langevin updates in two chains, we need only use the same
random normal vector in step (1), and the same random number for the
acceptance decision in step (5).  This method is quite effective.
Figure~\ref{fig-lang1} shows the result of using it to sample from the
nine-dimensional normal distribution described in
Section~\ref{sec-rgmvnex}.  The distance between states in the two
chains declines exponentially fast, reaching quite a small value in
the same time as is needed for one of the chains to reach a point
nearly independent of its initial state.  

The good coupling behaviour of Langevin updates for multivariate
normal distributions is easily explicable.  For a zero-mean normal
distribution with covariance $\Sigma$, the uncorrected update of
equation~(\ref{eq-ulang}) is
\beq
  x_{t+1} & = & 
   x_t \ -\ (\epsilon^2/2) \Sigma^{-1} x_t\ +\ \epsilon u_t
\eeq
The separation between two chains, $x$ and $x'$, will therefore 
change as follows:
\beq
  x_{t+1}-x'_{t+1} & = & [I - (\epsilon^2/2)\Sigma^{-1}]\,(x_t - x'_t)
\eeq
If the eigenvalues of $\Sigma^{-1}$ are $\lambda_i$, the eigenvalues
of $I-(\epsilon^2/2)\Sigma^{-1}$ will be $1-(\epsilon^2/2)\lambda_i$.
These eigenvalues will all have absolute value less than one if 
$\epsilon$ is less than $2/\sqrt{\lambda_{\mbox{\small max}}}$, where 
$\lambda_{\mbox{\small max}}$ is the largest eigenvalue of $\Sigma^{-1}$.
Coupled uncorrected Langevin chains will therefore approach each other
rapidly as long as $\epsilon$ is chosen to be somewhat smaller than this.  
For the corrected Langevin method, this good coupling behaviour might
be disturbed by rejections that occur in one chain but not the other.
If the rejection rate is low, however, this will not be a problem.
The example above used a stepsize of $\epsilon=0.08$, substantially
less than $2/\sqrt{\lambda_{\mbox{\small max}}} = 0.14$.  The
rejection rate was 14\%.

More generally, Langevin updates couple well when applied to 
any distribution over $R^n$ for which the log of the
density function is strictly concave.  Consider first a univariate
distribution of this sort.  Suppose the current states of two coupled
chains sampling from this distribution are $x_t$ and $x'_t$, and suppose
without loss of generality that $x'_t<x_t$.  The
log-concavity of the density then guarantees that $\nabla E(x'_t) <
\nabla E(x_t)$, from which it follows that, for an uncorrected Langevin
update,
\beq
  x_{t+1}-x'_{t+1}\ =\ 
   x_t-x'_t\ -\ (\epsilon^2/2)\,(\nabla E(x_t) - \nabla E(x'_t))
   \ <\ x_t-x'_t 
\eeq
It is possible for this update to overshoot, with $x_{t+1}-x'_{t+1}$
being negative, and $|x_{t+1}-x'_{t+1}|$ possibly being
greater than $|x_t-x'_t|$.  However, if the density is smooth, 
and $\epsilon$ is sufficiently small, $E(x)$ will be well-approximated
by a quadratic function in the vicinity of $x_t$ and $x'_t$ when these
points are
close together.  The chains will then tend to approach for the
same reason as they do when sampling from a normal distribution,
as discussed above.
Corrected Langevin updates will also tend to produce good coupling
as long as $\epsilon$ is small enough that the rejection rate is low.

This argument extends to multivariate log-concave distributions as
long as the stepsize, $\epsilon$, is small.  To see this, consider the
line passing through the current states of the two chains, $x$ and
$x'$.  The probability density along this line will be log-concave if
the joint density is log-concave.  If the Langevin update moved only
along this line, according to the projections on it of $\nabla E(x)$
and $\nabla E(x')$, the argument above would show that the distance
between the states of the two chains would decrease.  Generally, of
course, the states will also move in directions orthogonal to this
line.  However, to first order, small movements in these orthogonal
directions do not change the distance between $x$ and $x'$.  The
states of the chain will therefore tend to approach each other when
$\epsilon$ is sufficiently small.

The states of two coupled Langevin chains need not always approach
each other when the distribution is not log concave.  However, we
might nevertheless hope that reasonably good coupling behaviour will
result as long as the chains spend enough time in regions of the state
space where the density is log concave.  Further investigation is
needed to understand when this is actually true.

The efficiency of Langevin sampling can usually be improved by modifying
step~(1) of the procedure shown above as follows (Horowitz 1991; Neal
1996a, Section 3.5.3):\vspace*{-5pt}
\begin{enumerate}
\item[1b.] Change $p$ to $\alpha p + (1-\alpha^2)^{1/2} n$, where $n$
           is an independent draw from the normal distribution
           with mean zero and covariance $I$.\vspace*{-5pt}
\end{enumerate}
Setting $\alpha$ to zero produces the standard Langevin method.  When
a value for $\alpha$ close to one is used, the chain tends to proceed
in approximately the same direction for many iterations (provided the
rejection rate is small).  This ``persistent'' form of the Langevin
method suppresses the random walk behaviour
of standard Langevin updates, thereby improving the efficiency with which the
space is explored.

\begin{figure}[t]

\centerline{\psfig{file=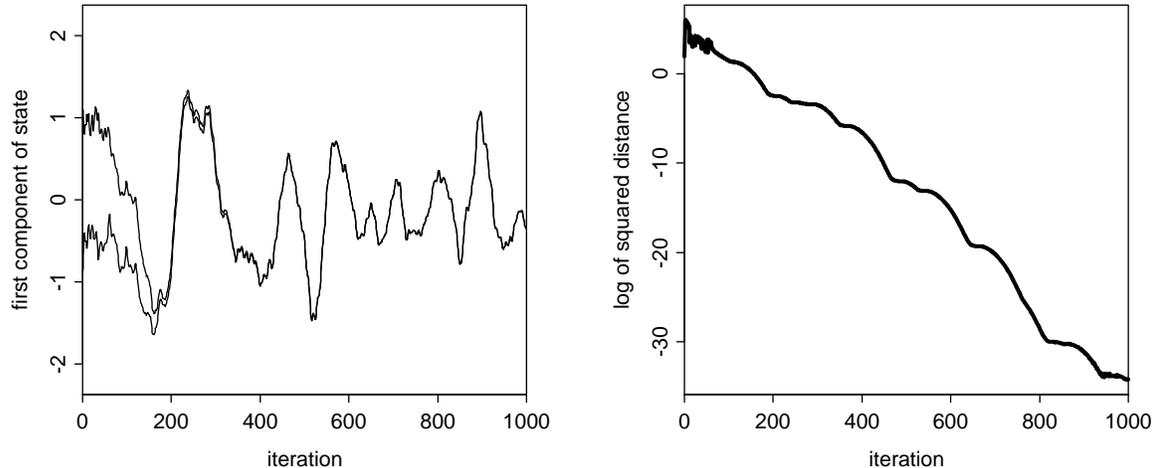}}

\vspace*{-20pt}

\caption[]{Sampling from a multivariate normal distribution using
coupled Langevin chains with persistence ($\alpha=0.95$,
$\epsilon=0.04$).  The plot on the left shows the first component of
state for the two coupled chains.  The plot on the right show the log
of the squared Euclidean distance between the states of the two chains
(the momentum, $p$, is included in the state).}\label{fig-lang2}

\end{figure}

Figure~\ref{fig-lang2} shows the performance of Langevin with
persistence on the nine-dimensional multivariate normal distribution,
using $\alpha=0.95$ and $\epsilon=0.04$.  This value for $\alpha$
suppresses random walks for roughly 20 iterations, until the momentum
has been largely randomized by the updates of step (1b).  A smaller
stepsize was used here than for the standard Langevin updates so that
the rejection rate would be small, avoiding the undesirable reversals
of direction that occur when the proposed state is rejected in
step~(5).  The plot on the left clearly shows that the persistent form
of Langevin samples more efficiently than standard Langevin (compare
with the left plot in Figure~\ref{fig-lang1}).  Furthermore, the
states of the coupled chains approach each other very closely, in time
comparable to that needed for one of the chains to move to a roughly
independent point.

\subsection{Coupled Gibbs sampling updates}\label{sec-cgibbs}\vspace*{-6pt}

Gibbs sampling updates can be effectively coupled when random variates
from the successive full conditional distributions are generated by
inversion of the cumulative distribution function, or by a procedure
with an equivalent coupling effect.  This method has used by Johnson
(1996) and by Pinto and Neal (2001).

In detail, we implement Gibbs sampling for the $i$th component by replacing
$x_i$ by 
\[
   F^{-1}(u_i\,|\,x_1,\ldots,x_{i-1},x_{i+1},x_n)
\]
where $u_i$ is a random variate uniformly distributed on $(0,1)$, and
$F^{-1}$ is the inverse of the cumulative distribution function for
the conditional distribution of $x_i$ given the current values of the
other components.  This method for generating random variates with
a given distribution is efficient for a few distributions (eg, the
exponential and the Cauchy), but it would not be the preferred method 
for most distributions if coupling were not an issue.  Fortunately, if 
the conditional distributions for different chains differ only by
translation and scaling, as will be the case when sampling from
a multivariate normal, an equivalent coupling effect can be obtained by
transforming a random variate with location parameter zero and 
scale parameter one, generated by any convenient method, to the 
appropriate location and scale in each chain.  

\begin{figure}[t]

\centerline{\psfig{file=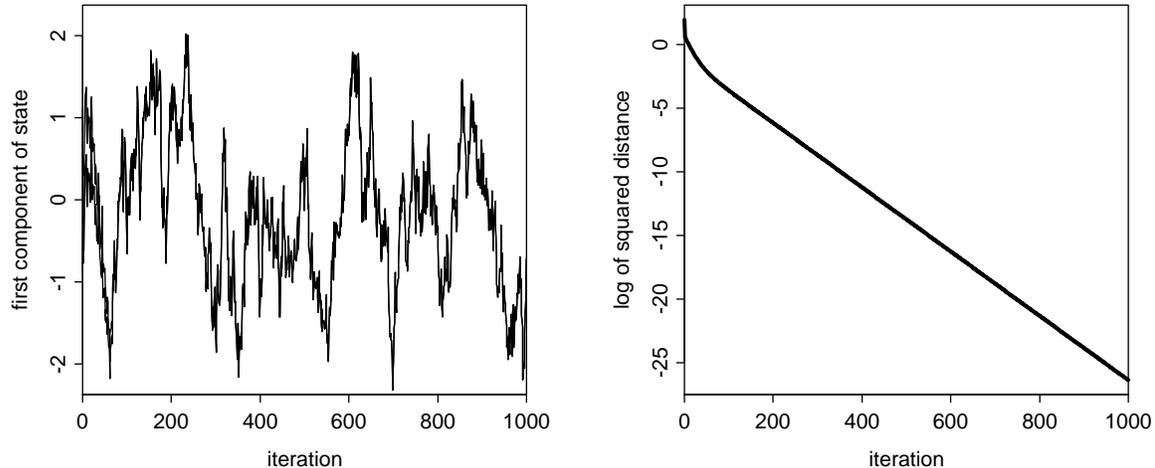}}

\vspace*{-20pt}

\caption[]{Coupled Gibbs sampling for a multivariate normal distribution.
The plot on the left shows the first component of
state for the two coupled chains.  The plot on the right show the log
of the squared Euclidean distance between the states of the two 
chains.}\label{fig-gs1}

\end{figure}

Figure~\ref{fig-gs1} demonstrates the effectiveness of this coupling
method for the example nine-dimensional normal distribution.  Pinto
and Neal (2001) found that this coupling scheme is also very effective
for some non-normal distributions.  Unfortunately, although the
adaptive rejection sampling method of Gilks and Wild (1992) allows
Gibbs sampling to be done for any log-concave distribution, it is not
easy to see how to modify adaptive rejection sampling to produce good
coupling.  This limits the contexts in which coupled Gibbs sampling
is possible.

\section{A logistic regression example}\label{sec-lr}\vspace*{-10pt}

To see whether circular coupling works in practice for typical
Bayesian inference problems of moderate complexity, I have tested it
using simulated data for a polytomous logistic regression model with a
hierarchical prior.

The problem concerns
four real-valued predictor variables and one three-way class variable,
taking values of 1, 2, or 3.
The conditional probability of the class variable in case $i$, 
written $c_i$, given the predictor variables, $x_{i1},\ldots,x_{i4}$,
is modeled as follows:
\beq
  P(c_i=k\ |\ x_{i1},\ldots,x_{i4}) & = & \exp(z_{ik}) \Big/ 
                  \sum_{k'=1}^3 \exp(z_{ik'})
\eeq
where
\beq
 z_{ik} & = & b_{0k}\ +\ \sum_{j=1}^4 b_{jk} x_{ij}
\eeq

The model parameters $b_{jk}$ are redundant, since adding a constant
to $b_{jk}$ for all $k$ does not change the probabilities of the
$c_i$.  This redundancy is retained in order to permit easy
specification of a prior that is symmetric with respect to the three
classes.  A hierarchical prior is used, in which the prior
distribution of the $b_{jk}$ depends on hyperparameters $\tau_j$
associated with the predictor variables, whose prior distribution is
in turn controlled by a top-level hyperparameter, $\tau_*$.  The
prior specifications are as follows (with the only 
dependencies present being those explicitly shown):
\beq
   b_{0k} & \sim & N(0,1),\ \ \mbox{for $k=1,\ldots,3$} \\[4pt]
   b_{jk}\,|\,\tau_j 
   &\sim & N(0,1/\tau_j),\ \ \mbox{for $j=1,\ldots,4$ and $k=1,\ldots,3$}\\[4pt]
   \tau_j\,|\,\tau_* & \sim & \mbox{Exp}(\tau_*),\ \ \mbox{for $j=1,\ldots,4$}
   \\[4pt]
   \tau_* & \sim & \mbox{Exp}(1)
\eeq

\begin{figure}[p]

\centerline{\psfig{file=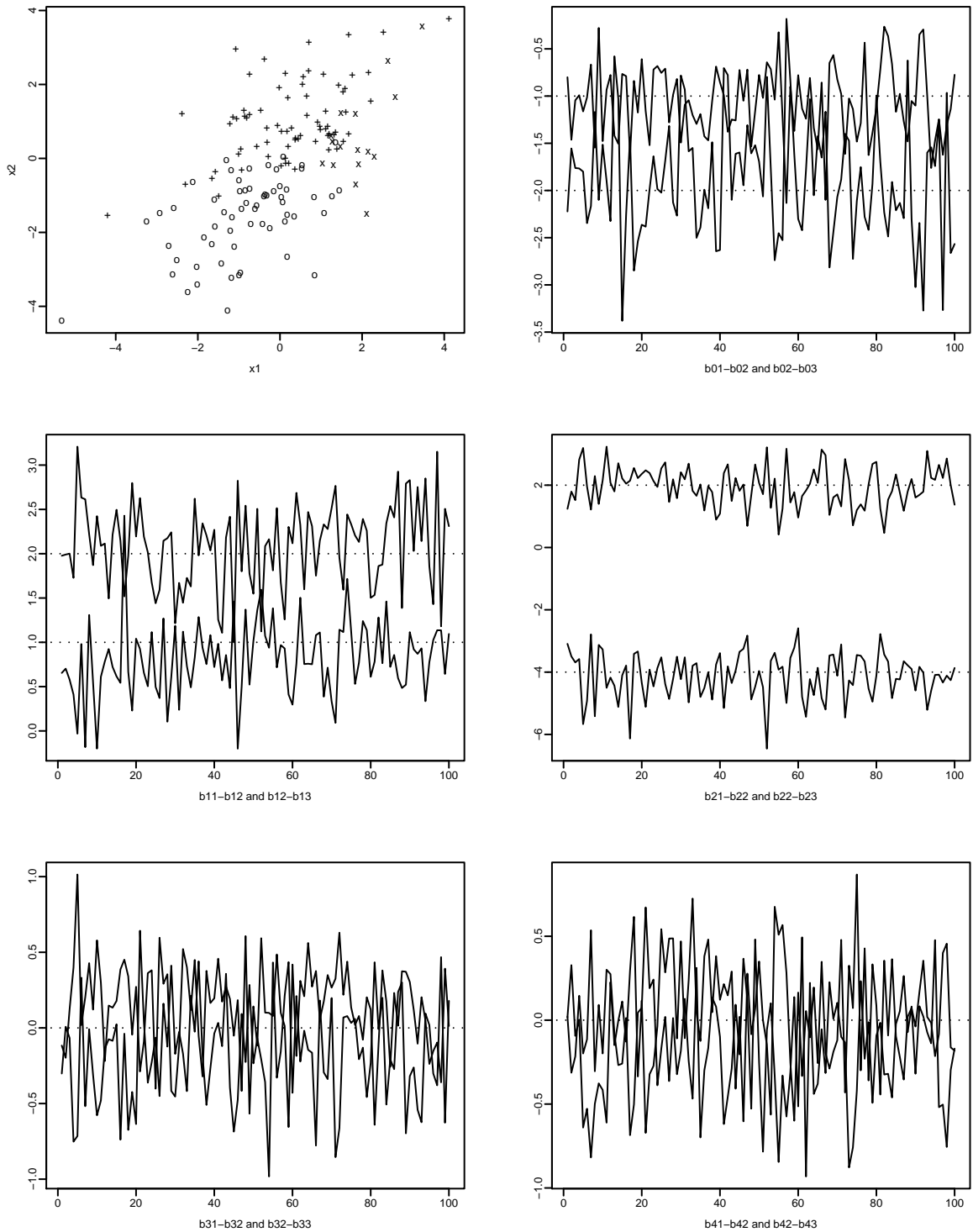}}

\vspace*{-20pt}

\caption[]{The top left plot shows the 150 simulated data points, plotted 
according to $x_{i1}$ and $x_{i2}$, with $c_i$ indicated by the symbol 
(0=x, 1=o, 2=+). The remaining plots show the differences $b_{j1}-b_{j2}$
and $b_{j2}-b_{j3}$ for $j=0,\ldots,4$, plotted for the 100 states obtained
from the circularly coupled Markov chain.  The dashed lines indicate the
true values for these parameter differences.}\label{fig-lr1} 

\end{figure}

A data set of size 150 was simulated in which the four predictor
variables had a multivariate normal distribution, each with mean zero,
variance 2, and correlation 1/2 with each of the other three predictor
variables.  The class variable was simulated using the values chosen 
for the predictor variables along with the following values for the parameters:
\beq
   \begin{array}{lllll}
   b_{01} = -2 ~&~ b_{11} = 3 ~&~ b_{21} =  0 ~&~ b_{31} = 0 ~&~ b_{41} = 0 \\
   b_{02} =  0 ~&~ b_{12} = 1 ~&~ b_{22} = -2 ~&~ b_{32} = 0 ~&~ b_{42} = 0 \\
   b_{03} =  1 ~&~ b_{13} = 0 ~&~ b_{23} =  2 ~&~ b_{33} = 0 ~&~ b_{43} = 0
   \end{array}
\eeq
Notice that the last two predictor variables have no effect on the distribution
of the class variable.  The first two predictor variables and the class 
are plotted for the 150 simulated cases in the top left of Figure~\ref{fig-lr1}.

I sampled from the posterior distribution for this model and data set
using a hybrid strategy.  With the hyperparameters temporarily fixed,
the parameters, $b_{jk}$, were updated using the Langevin method with
persistence, along with occasional random-grid Metropolis updates.
The top-level hyperparameter, $\tau_*$, was updated using random-grid
Metropolis alone.  The lower-level hyperparameters, $\tau_j$, were
updated by Gibbs sampling alone.  Random-grid updates were not needed
to produce exact coalescence for these hyperparameters because
conditional on $\tau_*$ and the $b_{jk}$, the $\tau_j$ are independent
of each other, and hence will coalesce exactly with one Gibbs sampling
update once $\tau_*$ and the $b_{jk}$ have coalesced exactly.

\pagebreak

In detail, one complete iteration of the Markov chain consisted of
the following updates:\vspace*{-10pt}
\begin{enumerate}
\item[1)] 10 repetitions of the following:\vspace*{-4pt}
\begin{enumerate}
\item[a)] 10 Langevin updates with stepsize $\epsilon=0.05$ and with
          persistence $\alpha=0.97$.
\item[b)] 25 random-grid Metropolis updates for $\log(\tau_*)$, using a proposal
          with $w=0.1$.
\item[c)] One Gibbs sampling update for each of the $\tau_j$.\vspace*{-4pt}
\end{enumerate}
\item[2)] A random-grid Metropolis update for all the $b_{jk}$ simultaneously,
      using a proposal with $w=0.01$.\vspace*{-4pt}
\item[3)] A random-grid Metropolis update for $\log(\tau_*)$, using a proposal
      with $w=0.1$.\vspace*{-4pt}
\item[4)] One Gibbs sampling update for each of the $\tau_j$.\vspace*{-4pt}
\item[5)] A replacement of the momentum variables by values drawn independently
      from the standard normal distribution.\vspace*{-10pt}
\end{enumerate}      

The parallel simulation algorithm of Section~\ref{sec-par} with
$N=100$ and $r=10$ was applied, although only one physical processor
was used.  Ten initial states were sampled from the prior distribution
to begin each of the ten segments of the simulation.  Each segment
continued from this initial state for ten iterations of the above
steps.  Each segment was then restarted from the final state of the
preceding segment, and re-simulated until it coalesced with the
previously simulated chain, or the ten iterations were completed.
This re-simulation was performed again whenever the start state of the
segment changed.

Coalescence was reasonably quick.  In the first re-simulation stage,
one of the ten segments coalesced with the previous simulation after 9
iterations; the others continued until all ten iterations were done.
In the second re-simulation stage (involving only nine segments, since
one segment's start state was unchanged), six of the segments
coalesced with the previous simulation (after 1, 1, 1, 5, 9, and 9
iterations).  The third re-simulation stage involved only three
segments, which all coalesced (after 1, 6, and 6 iterations),
completing the process.  A total of 268 iterations were simulated,
requiring 16 seconds of computation time on a 1.7 GHz Pentium 4
processor.

\begin{figure}[t]

\centerline{\psfig{file=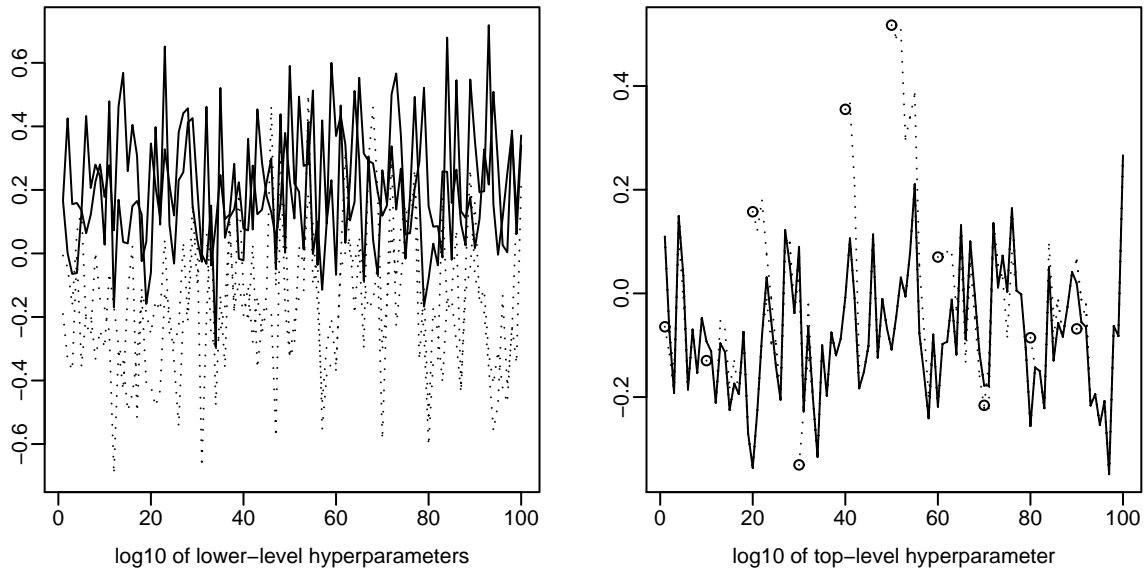}}

\vspace*{-20pt}

\caption[]{Hyperparmeter values from the circularly coupled chain.
The left plot shows the lower-level hyperameters, in the form of
$\log_{10}(1/\sqrt{\tau})$.  The hyperparameters for the first two
predictor variables are shown with solid lines, those for the last two
with dotted lines.  The right plot shows the values of the top-level
hyperparameter, in the form of $\log_{10}(1/\sqrt{\tau_*})$.  The solid
line gives the values for this hyperparameter at the states of the
final wrapped-around chain.  The dotted lines show the values of this
hyperparameter in the chains simulated from the ten initial states,
indicated by circles.}\label{fig-lr2}

\end{figure}

Simulation traces for the relevant differences in the parameter values
are shown in Figure~\ref{fig-lr1}, for the 100 states of the final
wrapped-around chain (the state at time 0 is omitted, since it is
identical to the state at time 100).  The left plot in
Figure~\ref{fig-lr2} shows a similar trace of the four lower-level
hyperparameters.  Note that the hyperparameters controlling the
magnitudes of the parameters for the two irrelevant inputs have taken
on values that shrink the distribution of the regression coefficients
for these variables to be close to zero.

Finally, the right plot in Figure~\ref{fig-lr2} shows a trace of the
top-level hyperparameter for the wrapped-around chain, along with
traces starting at the ten initial states, showing that the chains
started from these states coalesce rapidly with the wrapped-around
chain.  This rapid coalescence from a variety of starting states
provided evidence (though not a certain guarantee) that the
assumptions required for the states of the wrapped-around chain to
have close to the desired distribution are satisfied.

\section{Discussion}\label{sec-disc}\vspace*{-10pt}

The circular coupling procedure has been shown here to produce states
with approximately the correct distribution provided certain
conditions regarding coalescence times are satisfied.  In practice,
we will usually not know with certainty whether these conditions hold,
but diagnostic tests can provide useful evidence of this.

If we did have a theoretical proof that the required conditions hold,
the benefits of circular coupling would be modest.  These conditions
would also suffice to show that the distribution of the last state of
an ordinary Markov chain simulation for $N$ time steps comes from
close to the equilibrium distribution of the chain.  We could
therefore obtain a sample of $N$ (dependent) points from approximately
the correct distribution by simply continuing the simulation of this
chain for another $N$ iterations.  The circular coupling procedure
would save at most a factor of two in computation time compared to
this alternative, less if coalescence of the wrapped-around chain did
not occur quickly.

The primary reason why circular coupling is of interest is that it may
provide a way of diagnosing convergence and of discarding a
``burn-in'' period that is more automatic than current methods,
allowing Markov chain Monte Carlo methods to be used on a more routine
basis.

We can compare circular coupling with two contrasting strategies for
Markov chain Monte Carlo that have been advocated by Geyer (1992) and
by Gelman and Rubin (1992).  Geyer favours simulating a single chain
for as long as possible, in order to maximize the chances that this
time is long enough for a good approximation to the equilibrium
distribution to be reached.  Gelman and Rubin advocate that the
available time be used instead to simulate a moderate number of
chains, which consequently are each run for a shorter time, using
starting states chosen from an ``overdispersed'' distribution.  This
allows one to recognize that there is a problem if chains started in
different states behave differently.

The benefits of both of these schemes are obtained using a single,
long circularly-coupled chain, together with a moderate number of
auxiliary diagnostic chains, whose starting points are chosen from an
overdispersed distribution.  There is no need to perform more than one
circularly-coupled simulation, provided the auxiliary diagnostic
chains are found to coalesce in much less time than the length of the
run.  Since the auxiliary chains are started from independent points,
and each uses initially a different segment of the random number
sequence, they provide independent checks on the convergence of the
chain, just as completely separate simulations would.

The strategy of using a single circularly-coupled run has the
disadvantage that the time required is not completely predictable,
since the coalescence time is not known in advance.  If coalescence is
rapid, the run may finish in less than the time available, and one
might wish to use the remaining time to simulate further states, in
order to obtain more accurate estimates.  On the other hand, if
coalescence turns out to be slow, a longer run may be necessary in
order to be sure that the correct distribution has been reached.  It
should be possible to extend a circularly-coupled simulation run to a
greater length, redoing only the very beginning, where the
wrapped-around starting state will change once the chain is run for
longer.  Further work is needed to work out the details of such a
procedure, and to investigate whether letting the run vary in length
might introduce bias.

Another advantage of diagnosing convergence based on coupling is that
it looks at the entire state.  Informal convergence diagnostics are
usually based instead on a small number of low-dimensional functions
of state.  Other methods for diagnosing convergence based on the whole
state, such as that of Ritter and Tanner (1992), are likely to break
down when the state is of high dimension.  In contrast, at least in
some cases, such as sampling from a multivariate normal using the
Langevin method (see Section~\ref{sec-clang}), circular coupling can
work well even for a very high dimensional distribution.

Assuming that we have determined to our satisfaction that the chain
has converged, we still have the problem of discarding an initial
``burn-in'' segment.  Wrapping the chain around to produce a
circularly-coupled chain solves this problem without introducing any
substantial bias, under the assumptions of the theorem in
Section~\ref{sec-proof}, whereas other methods for deciding how much
of the chain to discard can introduce bias into the final result, as
discussed by Cowles, \textit{et al} (1999).

A further minor advantage of circular coupling is that assessing the
accuracy of estimates by accounting for the autocorrelation along the
chain is slightly easier when the chain is circular, because time
series techniques work more cleanly with circular time series than
with time series having ends.  For instance, the need to ``taper'' the
ends when doing spectral analysis is eliminated.  

The ability of circular coupling to exploit parallel computation may
also be useful in practice.  The parallel procedure of
Section~\ref{sec-par} can be routinely applied whenever circular
coupling can be done at all.  If one is simulating a single, long
circular chain, the only other possibility for exploiting parallelism
would be within the computations needed to perform the individual
transitions.  Though there will typically be much scope for this, it
does require special programming for the particular chain being
simulated.

To obtain these benefits, effective coupling schemes will need to be
developed, which can be applied to a wide range of distributions with
little or no need for problem-specific tailoring.  The strategy of
combining Langevin and Gibbs sampling updates with random-grid
Metropolis updates has been shown here to work well for a non-trivial
Bayesian inference problem, but further theoretical and empirical work
is needed to characterize the range of problems for which these
methods will work well.  More sophisticated Markov chain methods,
notably ``hybrid Monte Carlo'', are needed for the most difficult
problems, such as Bayesian inference for neural network models (Neal
1996a).  In preliminary work, I have found that a combination of
hybrid Monte Carlo and random-grid Metropolis leads to rapid
coalescence for some problems, but for others, such as inference for
complex neural network models, coalescence is difficult to achieve.
Solving this problem may require some additional innovation, such as,
perhaps, the use of ``tempering'' methods (Neal 1996b) to produce a
simplified distribution in which chains can more easily be brought
together.

\section*{Acknowledgements}\vspace*{-10pt}

I thank Jeffrey Rosenthal for inspiration to examine the possibility
of parallel simulation of Markov chains and for helpful discussions.
This research was supported by the Natural Sciences and Engineering
Research Council of Canada, and by the Institute for Robotics and
Intelligent Systems.

\section*{References}\vspace*{-10pt}

\leftmargini 0.2in
\labelsep 0in

\begin{description}
\itemsep 2pt

\item Brooks, S.~P.\ and Roberts, G.~O.\ (1998) ``Convergence assessment
      techniques for Markov chain Monte Carlo'', \textit{Statistics and
      Computing}, vol.~8, pp.~319-335.

\item Chen, M.-H.,\ Shao, Q.-M., and Ibrahim, J.~G.\ (2000)
      \textit{Monte Carlo Methods in Bayesian Computation},
      Springer-Verlag.

\item Cowles, M.~K.\ and Carlin, B.~P.\ (1996) ``Markov chain Monte Carlo
      convergence diagnostics: a comparative study'', \textit{Journal of
      the American Statistical Association}, vol.~91, pp.~883-904.

\item Cowles, M.~K., Roberts, G.~O., and Rosenthal, J.~S.\ (1999) ``Possible 
      biases induced by MCMC convergence diagnostics'',  \textit{Journal of
      Statistical Computation and Simulation}, vol.~ 64, pp.~87-104.

\item
  Duane, S., Kennedy, A.~D., Pendleton, B.~J., and Roweth, D.\ (1987)
  ``Hybrid Monte Carlo'', {\em Physics Letters B}, vol.~195, pp.~216-222.

\item Fill, J.~A.\ (1998) ``An interruptible algorithm for perfect sampling
      via Markov chains'', \textit{Annals of Applied Probability} vol.~8,
      pp.~131-162.

\item Frenkel, D.\ and Smit, B.\ (1996) \textit{Understanding
      Molecular Simulation: From Algorithms to Applications}, 
      San Diego: Academic Press.

\item Gelman, A.\ and Rubin, D.~B.\ (1992) ``Inference from iterative
      simulation using multiple sequences'' (with discussion),
      \textit{Statistical Science}, vol.~7, pp.~457-472 
      (discussion pp.~483-511).

\item Geyer, C.~J.\ (1992) ``Practical Markov chain Monte Carlo'',
      \textit{Statistical Science}, vol.~7, pp.~473-483
      (discussion pp.~483-511).

\item Gilks, W.~R., Richardson, S., and Spiegelhalter, D.~J.\ (1996)
      \textit{Markov Chain Monte Carlo in Practice}, London: Chapman and Hall.

\item
  Gilks, W.~R.\ and Wild, P.\ (1992) ``Adaptive rejection sampling
  for Gibbs sampling'', \textit{Applied Statistics}, vol.~41, pp.~337-348.

\item Green, P.~J.\ and Murdoch, D.~J.\ (1998) ``Exact sampling for 
      Bayesian inference: towards general purpose algorithms'' 
      (with discussion), in J.~M.~Bernardo, \textit{et al} (editors), 
      \textit{Bayesian Statistics 6}, Oxford: Clarendon Press, pp.~301-321.

\item
  Hastings, W.~K.\ (1970) ``Monte Carlo sampling methods using Markov chains 
  and their applications'', {\em Biometrika}, vol.~57, pp.~97-109.

\item
  Horowitz, A.~M.\ (1991) ``A generalized guided Monte Carlo algorithm'',
  {\em Physics Letters B}, vol.~268, pp.~247-252.

\item Johnson, V.~E.\ (1996) ``Studying convergence of Markov chain 
      Monte Carlo algorithms using coupled sample paths'', \textit{Journal
      of the American Statistical Association}, vol.~91, pp.~154-166.

\item Johnson, V.~E.\ (1998) ``A Coupling-Regeneration Scheme for Diagnosing 
      Convergence in Markov Chain Monte Carlo Algorithms'',
      \textit{Journal of the American Statistical Association}.

\item Lindvall, T.\ (1992) \textit{Lectures on the Coupling Method}, 
      New York: Wiley.

\item Mengersen, K.~L., Robert, C.~P., and Guihenneuc-Jouyaux, C.\ (1999)
      ``MCMC convergence diagnostics: a reviewww'' (with discussion), in 
      J.~M.~Bernardo, \textit{et al} (editors), \textit{Bayesian Statistics 6},
      Oxford: Clarendon Press, pp.~415-440.

\item Neal, R.~M.\ (1996a) \textit{Bayesian Learning for Neural Networks},
      Lecture Notes in Statistics No.\ 118, New York: Springer-Verlag.

\item Neal, R.~M.\ (1996b) ``Sampling from multimodal distributions using
  tempered transitions'', {\em Statistics and Computing}, vol.~6, pp.~353-366.

\item Pinto, R.~L.\ and Neal, R.~M.\ (2001) ``Improving Markov chain Monte 
      Carlo estimators by coupling to an approximating chain'', Technical 
      Report No. 0101, Dept. of Statistics, University of Toronto, 13 pages.

\item Propp, J.~G.\ and Wilson, D.~B.\ (1996) ``Exact sampling with coupled
      Markov chains and applications to statistical mechanics'', \textit{Random
      Structures and Algorithms}, vol.~9, pp.~223-252.


\item
  Ritter, C.\ and Tanner, M.~A.\ (1992) ``Facilitating the Gibbs sampler:
  The Gibbs stopper and the Griddy-Gibbs sampler'', {\em Journal of the
  American Statistical Association}, vol.~87, pp.~861-868.

\item Roberts, G.~O., Gelman, A., and Gilks, W.~R.\ (1997) 
      ``Weak convergence and optimal scaling of random walk Metropolis
      algorithms'', \textit{Annals of Applied Probability}, vol.~7, 
      pp.~110-120.

\item Rosenthal, J.~S.\ (1995a) ``Minorization Conditions and
      Convergence Rates for Markov Chain Monte Carlo'', \textit{Journal of
      the American Statistical Association}, vol.~90, pp.~558-566.

\item Rosenthal, J.~S.\ (1995b) ``Convergence rates of Markov chains'',
      \textit{SIAM Review}, vol.~37, pp.~387-405.

\item Rosenthal, J.~S.\ (1999) ``Parallel computing and Monte Carlo
      algorithms'', Technical Report No.~9902, Dept.\ of Statistics,
      University of Toronto.

\item
  Rossky, P.~J., Doll, J.~D., and Friedman, H.~L.\ (1978) ``Brownian
  dynamics as smart Monte Carlo simulation'', {\em Journal of Chemical
  Physics}, vol.~69, pp.~4628-4633.

\item
  Schmeiser, B.\ and Chen, M.-H.\ (1991) ``On random-direction Monte Carlo
  sampling for evaluating multidimensional integrals'', Working Paper 
  SMS91-1, Dept. of Statistics, Purdue University.

\item Sinclair, A.\ (1993) \textit{Algorithms for Random Generation and 
      Counting: A Markov Chain Approach}, Boston: Birkh\"{a}user.

\end{description}
\end{document}